\documentclass[fleqn,usenatbib]{mnras}

\usepackage{newtxtext,newtxmath}

\usepackage[T1]{fontenc}
\usepackage{ae,aecompl}

\usepackage{graphicx}	
\usepackage{amsmath}	
\usepackage{amssymb}	

\newcommand{\ramses}{\textsc{ramses}}
\newcommand{\ramsesrt}{\textsc{ramses-rt}}

\newcommand{\beq}{\begin{equation}}
\newcommand{\eeq}{\end{equation}}
\newcommand{\bt}{\begin{table}}
\newcommand{\et}{\end{table}}

\newcommand{\hei}    {{\rm{He\textsc{i}}}}
\newcommand{\heii}   {\rm{He\textsc{ii}}}
\newcommand{\heiii}   {\rm{He\textsc{iii}}}
\newcommand{\hi}    {{\rm{H\textsc{i}}}}
\newcommand{\hii}   {\rm{H\textsc{ii}}}
\newcommand {\htwo}{\rm{H$_2$}}

\newcommand{\htsub} {\rm{H \scriptscriptstyle 2}}

\newcommand{\xhi} {x_{\rm{H \scriptscriptstyle I}}}
\newcommand{\xhii} {x_{\rm{H \scriptscriptstyle II}}}
\newcommand{\xhtwo} {x_{\rm{H \scriptscriptstyle 2}}}

\newcommand{\cci} {{\rm{cm}}$^{-3}$}   
\newcommand{\si} {\rm{s}$^{-1}$}       
\newcommand{\kms} {\rm{kms}$^{-1}$}       
\newcommand{\msun}{\rm{M$_{\sun}$}}

\title[Molecular galaxy census]{The complete census of molecular hydrogen in a simulated disc galaxy}

\author[Sarah Nickerson et al.]{
Sarah Nickerson,$^{1}$\thanks{E-mail: snickers@physik.uzh.ch}
Romain Teyssier,$^{1}$
Joakim Rosdahl$^{2}$
\\
$^{1}$Institute for Computational Science, University of Z{\"u}rich, Winterthurerstrasse 190, CH-8057 Z{\"u}rich, Switzerland\\
$^{2}$Centre for Astronomy Research, University of Lyon, 9 avenue Charles Andr{\'e}, 69230 Saint-Genis-Laval, France\\
}

\date{Accepted XXX. Received YYY; in original form ZZZ}

\pubyear{2018}

\begin{document}
\label{firstpage}
\pagerange{\pageref{firstpage}--\pageref{lastpage}}
\maketitle

\begin{abstract}
\label{sec:abs}
We present a multi-scale analysis of molecular hydrogen in a Milky Way-like simulated galaxy. Our census covers the gas content of the entire disc, to radial profiles and the Kennicutt-Schmidt relation, to a study of its molecular clouds, and finally down to a cell-by-cell analysis of the gas phases. Where observations are available we find agreement. A significant fraction of the \htwo\ gas is in low-density regions mixed with atomic hydrogen and would therefore be difficult to observe. We use the molecular addition to \ramsesrt, an adaptive mesh refinement grid code with the hydrodynamics coupled to moment-based radiative transfer. Three resolutions of the same galaxy detail the effects it has on \htwo\ formation, with grid cells sized 97, 24, and 6.1 pc. Only the highest resolution yields gas densities high enough to host significant \htwo\ fractions, and resolution is therefore key to simulating \htwo. Apart our pieces of galactic analysis are disparate, but assembled they provide a cohesive portrait of \htwo\ in the interstellar medium. \htwo\ chemistry on the atomic scale is sufficient to generate its dynamics throughout an entire galaxy.
\end{abstract}

\begin{keywords}
galaxies: formation -- molecular processes -- radiative transfer -- methods: numerical.
\end{keywords}



\section{Introduction}
\label{sec:intro}

Molecular hydrogen (\htwo) is the most common molecule in the Universe \citep{Herbst2001}. It is critical to cooling the interstellar medium \citep{Glover2008}. Cold and dense giant molecular clouds (GMCs) are associated with young stars and star formation \citep{Blaauw1964,Werner1977,Blitz1980,Genzel1989}. Despite its importance, molecular hydrogen is challenging to  detect directly through observations \citep{Young1991}. CO is a secondary tracer of dense gas and is easily observed. \htwo\ abundance is inferred by a conversion factor from CO abundance, conventionally taken as a constant for the Milky Way \citep{Bolatto2013}, and in external galaxies it varies mainly as a function of metallicity \citep{Accurso2017}. Modern observations are pushing forward the boundaries of our understanding of \htwo\ to increased precision and more distant galaxies. \htwo\ is measured extensively for the Milky Way, the Local Group, and galaxies beyond, on scales that range from total gas content, to radially resolved profiles, and to individual giant molecular clouds.
 
The Kennicutt-Schmidt (KS) relation links the star formation rate (SFR) surface density to the gas surface density by a power law. \citet{Schmidt1959} infers this from analysis in the solar neighbourhood. Later, \citet{Kennicutt1989} successfully applies this relation to the overall surface densities of external galaxies. New measurements that resolve the structure of external galaxies and the \htwo\ within them \citep{Bigiel2008} find that the correlation is tighter between \htwo\ and SFR than total neutral hydrogen and SFR, as \citet{Leroy2013} confirm. The combined xCOLD GASS \citep{Saintonge2017} CO and xGASS \citep{Catinella2018} surveys of over a thousand galaxies find a weak correlation between the total molecular gas content and the SFR.

In an early study of Galactic GMCs, \citet{Larson1981} derives a relation in which the cloud's velocity dispersion scales with its size. \citet{Solomon1987} refine this relation with a larger catalogue, and \citet{Heyer2009} challenge this relation to show that the velocity dispersion not only depends on the size but also the surface density of the molecular cloud. They argue that earlier surveys lacked proper spatial resolution and mistook cloud surface density to be constant. \citet{Rice2016} with data from \citet{Dame2001} study GMCs in all regions of the Galaxy to show that the Larson relation holds throughout the Galaxy, but that the inner Galaxy tends to have more massive clouds. \citet{Miville-Deschenes2017} process this data differently and also provide a detailed analysis of cloud properties. New surveys \citep{Rosolowsky2007,Bolatto2008,Colombo2014,Leroy2017,Sun2018} create GMC catalogues beyond the Milky Way and discover a similar environment and relations, though \citet{Colombo2014} find the original Larson relation between velocity dispersion and cloud size to be weak.

The era of increasing precision for \htwo\ measurement requires a similar increase in sophistication for simulations. The very first galaxy simulations are dark matter only, but later editions include the hydrodynamical gas component and with it star formation and feedback from supernovae (SN). This sets the stage for describing the chemical species in the galactic gas.

Several codes use an equilibrium model to define the \htwo\ content, under the assumption that the chemistry in each volume element is in an equilibrium state determined by purely local variables. The most common formulation, in use by \citet{Kuhlen2012,Halle2013,Thompson2014}; and \citet{Hopkins2014}, is developed in \citet{Krumholz2008,Krumholz2011}; and \citet{McKee2010} (hereafter KMT) where the \htwo\ fraction is calculated from postulating an \hi-\htwo\ sphere in a homogeneous radiation field. \citet{Krumholz2013} updates this model for the molecular-poor regime. The equilibrium model of \citet{Robertson2008} is based on the photoionization code \textsc{cloudy} \citep{Ferland1998}, and \citet{Pelupessy2006} use a subgrid model of cloud populations. 

\citet{Gnedin2009} pioneers a non-equilibrium chemical network for \htwo, employing rate equations to track locally atomic and molecular hydrogen in their galaxy simulations with the \textsc{art} code \citep{Kravtsov1999}. \citet{Gnedin2011} expands this network, introducing ionized hydrogen and helium species. Many chemical networks for nearly as many codes follow. \citet{Christensen2012} and \citet{Tomassetti2014} adapt this method for \textsc{gasoline} \citep{Wadsley2004} and \ramses\ \citep{Teyssier2002} respectively. \citet{Baczynski2015} adapt the chemical network in \citet{Nelson1997}, \citet{Glover2007a}, \citet{Glover2007b}, and \citet{Glover2012} for \textsc{flash4} \citep{Fryxell2000,Dubey2008}; \citet{Richings2016} adapt the network in \citet{Richings2014a,Richings2014b} for \textsc{gadget3} \citep{Springel2005a}; \citet{Hu2016} also adapt the \citet{Glover2012} network but for \textsc{gadget3}; and \citet{Katz2017} follows \citet{Baczynski2015}'s adaptation for \ramsesrt\ \citep{Rosdahl2013}. \citet{Pallottini2017}, \citet{Capelo2018}, and \citet{Lupi2018} meld the \textsc{krome} \citep{Grassi2014} chemical network with \ramses, \textsc{gasoline2} \citep{Wadsley2017}, and \textsc{gizmo} \citep{Hopkins2015} respectively. 

A direct comparison between the KMT equilibrium model and non-equilibrium models shows that they diverge at low metallicities \citep{Krumholz2011}, that non-equilibrium models are capable of maintaining \htwo\ at lower densities \citep{Tomassetti2014},  and that the non-equilibrium models are clumpier and closer to the KS relation \citep{Pallottini2017}.

Non-equilibrium chemistry becomes even more powerful when coupled to the radiative transfer of the photons that dissociate and ionize the gas. Two radiative transfer methods are currently in use. \citet{Baczynski2015} employ ray tracing, which has the advantage of computing exact column densities. However, the computational cost of ray tracing is proportional to the number of radiation sources and this becomes less feasible in galaxy simulations filled with stars. Another option is to use a moment-based method in which the gas is treated as a fluid, and this is much more feasible for galaxy simulations. \citet{Gnedin2011}, \citet{Lupi2018}, and \ramsesrt\ \citep{Rosdahl2013} use this method. 

Many of the chemical networks follow similar models for the formation and destruction of \htwo, but two choices in how to model the subgrid physics differentiate them. 

The first is whether to include a clumping factor, the purpose of which is to account for unresolved dense structure in molecular clouds. Practically, this amounts to enhancing \htwo\ formation by some factor. The most commonly-used constant for the clumping factor \citep{Christensen2012,Katz2017,Capelo2018} is 10 from \citet{Gnedin2009} who reason that this is the ratio of average \htwo\ density to to cloud density \citep{McKee2007}. \citet{Micic2012} find that a constant clumping factor may lead to an over-prediction of \htwo\ in high-density regions. \citet{Capelo2018} compare the clumping factor of 10 to no clumping factor, and find that it does indeed enhance \htwo\ formation. Others employ a variable density-based clumping factor \citep{Tomassetti2014,Lupi2018}. \citet{Tomassetti2014} find their model with the constant clumping factor is closer to the KMT equilibrium model  compared to the variable clumping factor. Others opt out of the clumping factor completely \citep{Baczynski2015,Richings2016,Hu2016,Pallottini2017}.

The second choice is whether or not to link star formation explicitly to \htwo. Indeed, star formation is observationally correlated to \htwo\ \citep{McKee2007}, but this may be because both stars and \htwo\ form in dense cold environments and not because \htwo\ directly triggers star formation \citep{Glover2012}. \citet{Gnedin2011} argue that by setting the star formation rate proportional to the \htwo\ density, as opposed to total gas density as is traditional, they avoid possibly arbitrary density and temperature thresholds since \htwo\ naturally correlates with dense, cold gas. \citet{Christensen2012}, \citet{Tomassetti2014}, and \citet{Pallottini2017} follow suit. \citet{Richings2016}, \citet{Hu2016}, \citet{Katz2017}, \citet{Capelo2018}, and \citet{Lupi2018}, however, maintain a star formation rate proportional to total gas density. \citet{Hu2016} find in their simulations that star formation correlates with \hi-dominated cold gas better than \htwo, and see a significant quantity of warm, non-star-forming \htwo\ gas. \citet{Lupi2018} argue that the link is unnecessary because the KS relation between \htwo\ and SFR arises naturally without it.

A number of simulations that study molecular clouds in detail also include \htwo\ chemistry. \citet{Dobbs2008} use a non-equilibrium chemical network \citep{Bergin2004}, a constant dissociation rate for \htwo, and to calculate column density they adopt a constant length that represents the typical distance to a B0 star. This framework is also adopted by \citet{Khoperskov2013}, and \citet{Khoperskov2016} introduce ray tracing, while \citet{Duarte-Cabral2016} use radiative transfer post-processing.

In this paper we employ \ramsesrt\ \citep{Rosdahl2013}, which builds on the original \ramses\ adaptive mesh refinement (AMR) code \citep{Teyssier2002} by adding radiative transfer for photon groups coupled to the non-equilibrium chemistry of \hi, \hii, \hei, \heii, and \heiii. In \citet{Nickerson2018} we added the non-equilibrium chemistry of \htwo\ coupled to the radiative transfer. We tested this code in a number of idealized situations in order to successfully compare it against analytical solutions where they existed and other codes where they did not. Most notably we matched our results to  benchmark tests in photodissociation region (PDR) codes \citep{Roellig2007}. The only adjustable parameter was in our novel self-shielding model for \htwo, which took advantage of the radiative transfer. This was set by PDR scales smaller than those in use for galaxy simulations. We will not use a clumping factor in order to fully study the effects of resolution on \htwo\ content.  

In this paper we use our model for molecular, atomic, and ionized hydrogen coupled to radiative transfer in an isolated Milky Way-like disc galaxy at three resolutions. We aim to show that our model, calibrated on the chemical scale, gives rise to \htwo\ relations on the galactic scale without any tuning on this larger scale. The layout is as follows. Section \ref{sec:sim} describes the setup and physics of our simulations. We demonstrate our model's fidelity to realism on the galactic scale for the high resolution run and how resolution affects the molecular gas content of our galaxy in Section \ref{sec:gan}. In Section \ref{sec:mc} we employ a clump finder to identify molecular clouds to recover the Larson relation and cumulative mass profiles similar to observed molecular cloud populations. We discuss and summarize our findings in Section \ref{sec:conclu}.

\section{Simulation Setup}
\label{sec:sim}

We will first describe the initial conditions for our isolated Milky Way-like disc galaxy, before summarizing the physics and chemistry critical to this work.

\subsection{Initial conditions}
We study an isolated Milky Way-like galaxy originally generated for the AGORA comparison project \citep{Kim2014,Kim2016} at three resolutions, referred to as GHigh, GMed, and GLow. The square box width is 400 kpc. The dark matter halo initially follows a \citet{Navarro1997} profile, while the stellar and gas discs follow exponential profiles, and the stellar bulge follows a Hernquist profile \citep{Hernquist1990}. The parameters are summarized in Table \ref{tbl:galparams}; most apply to all three galaxies, while the particle number and cells size depend on the resolution. Each galaxy evolves to 800 Myr, more than enough time to settle into a semi-semi-steady state as discussed in Section \ref{sec:sfrhist}.

\begin{table*}
\begin{tabular}{crl|c|ccc}
\hline
\hline
Component & Parameter & & All  & GHigh & GMed & GLow\\
\hline
Dark Matter Halo &virial mass&$M_{200}$&$1.074\times10^{12}$\msun&-&-&-\\
&circular velocity&$v_{\text{c200}}$ & 150\kms&-&-&-\\
&virial radius&$r_{200}$&205.4 kpc&-&-&-\\
&concentration&$c$&10&-&-&-\\
&spin&$\lambda$&0.04&-&-&-\\
&particle number&$N_{\text{dm}}$&-&$10^7$&$10^6$&$10^5$\\
\hline
Stars&stellar disc mass&$M_{*\text{disc}}$&$4.297\times10^{10}$\msun&-&-&-\\
&disc radius&$r_{\text{disc}}$&3.432 kpc&-&-&-\\
&disc particle number&$N_{\text{*disc}}$&-&$10^7$&$10^6$&$10^5$\\
&bulge to disc ratio&$B/D$&0.1&-&-&-\\
&bulge particle number&$N_{\text{*bulge}}$&-&$1.25\times10^6$&$1.25\times10^5$&$1.25\times10^4$\\
&star formation threshold&$n_{*}$&-&300 \cci&50 \cci&8 \cci\\
&new particle mass&$m_{*}$&-&$10^3$ \msun&$10^4$ \msun&$10^5$ \msun\\
\hline
Gas Disc&disc gas fraction&$f_{\text{gas}}$&$20\%$&-&-&-\\
&initial metallicity&$Z_{\text{init}}$&1 Z$_{\sun}$&-&-&-\\
&min cell size&$\Delta x_{\text{min}}$&3125 pc&-&-&-\\
&max cell size&$\Delta x_{\text{max}}$&-&6.1 pc&24 pc&97 pc\\
\hline
\hline
\end{tabular}
\caption{Galaxy parameters for the three resolutions of galaxies, which are labelled: GHigh, GMed, and GLow. Parameters that apply to every galaxy fall under the ``All'' column, and those that differ fall under the column for each resolution. The dark matter halo parameters are $M_{200}$: halo virial mass as defined as the mass within a sphere 200 times more dense than the Universe's critical density, $v_{\text{c200}}$: circular velocity inside this sphere,  $r_{\text{200}}$: radius of this sphere, $c$: concentration parameter, $\lambda$: the spin parameter, and $N_{\text{dm}}$ the number of dark matter particles. The stellar component parameters are $M_{*\text{disc}}$: mass of the stellar disc, $r_{\text{disc}}$ radius of the disc, $N_{\text{*disc}}$ number of star particles in the disc, $B/D$: bulge to disc mass ratio, $N_{\text{*bulge}}$: number of star particles in the bulge, $n_{*}$: star formation critical density, and $m_{*}$: the mass of a new star particle. The gas disc properties are $f_{\text{gas}}$: gas fraction of the disc mass of gas and stars, $Z_{\text{init}}$: initial metallicity, $\Delta x_{\text{min}}$: minimum AMR resolution, and $\Delta x_{\text{max}}$: maximum AMR resolution.}
\label{tbl:galparams}
\end{table*}

\subsection{Ramses-RT}
We simulate our galaxies with \ramsesrt\ \citep{Rosdahl2013,Rosdahl2015a}, the radiative transfer extension of the hydrodynamics code \ramses\ \citep{Teyssier2002}. \ramses\ uses adaptive mesh refinement (AMR)  to model the gas with a second-order Godunov scheme and an N-body particle-mesh solver for dark matter and star particles. Our coarsest level is 7, and we refine up to levels 12, 14, and 16 for GLow, GMed, and GHigh respectively. Following a quasi-Lagrangian scheme, a cell is refined if it has ten or more dark matter and star particles older than 10 Myr, or if its combined gas and new stellar mass exceeds 10 $m_*$ (defined in Table \ref{tbl:galparams}).

\ramsesrt\ couples the hydrodynamics to the moment-based radiative transfer of photons using the M1 closure relation \citep{Levermore1984}. In particular, photons are split into discrete frequency groups depending on which species they dissociate or ionize, and their properties are integrated over the entire frequency range of the group. We use the reduced speed of light approximation \citep{Gnedin2001} in order to decrease computational time, specifically $c_r=c/200$ as in \citet{Rosdahl2015b} for galaxy simulations.

\subsection{Gas chemistry}
The chemical prescription we use here is the molecular hydrogen addition to \ramsesrt\ \citep{Nickerson2018}. We track the non-equilibrium chemistry of \htwo, \hi, \hii, \hei, \heii, and \heiii\ tied to the four dissociating and ionizing photon groups given in Table \ref{tbl:groups}. Details of the molecular chemistry are given in \citet{Nickerson2018} and the ionization chemistry in \citet{Rosdahl2013} but we summarize the processes here. 

\begin{table*}
\begin{tabular}{c|cc|c}
\hline
\hline
Group Number & Min $h\nu$ (eV) & Max $h\nu$ (eV)& Purpose\\
\hline
1 & 11.20 & 13.60 & \htwo\ dissociation \\
2 & 13.60 & 24.59 & \htwo\ and \hi\ ionization \\
3 & 24.59 & 54.42 & \htwo, \hi, and \hei\ ionization \\
4 & 54.42 & $\infty$ & \htwo, \hi, \hei, and \heii\ ionization \\
\hline
\hline
\end{tabular}
\caption{The photon group properties in our simulations, where $h\nu$ is the photon energy.}
\label{tbl:groups}
\end{table*}

Our \htwo\ model includes the formation of \htwo\ on dust \citep{Hollenbach1979,Jura1974,Gry2002,Habart2004}, as well as formation in the absence of dust via the gas phase \citep{McKee2010} and three-body collisions \citep{Forrey2013,Palla1983}, destruction by collision with \hi\ \citep{Dove1986} and with itself \citep{Martin1998}, photodissociation by the Lyman-Werner band (i.e., Group 1) \citep{Sternberg2014}, photoionization \citep{Abel1997}, and ionization by cosmic rays \citep{Indriolo2012,Gong2017,Glassgold1974}. In particular, we use a shortcut for the \htwo\ ionization. Instead of tracking the resultant species H$_2^+$, we treat \htwo\ ionization as a dissociation that creates two \hi\ atoms, under the assumption that an environment that ionizes \htwo\ is going to quickly ionize \hi. The \hi\ and He chemistry involves formation by recombination \citep{Hui1997}, collisional ionization \citep{Cen1992}, photoionsation \citep{Verner1996,Hui1997}, and cosmic ray ionization \citep{Indriolo2015,Gong2017,Glassgold1974,Glover2010}. We assume a hydrogen mass fraction of 0.76 and a helium mass fraction of 0.24

The gas is heated by photodissociation and ionization, the photoelectric effect \citep{Bakes1994,Wolfire2003}, UV pumping from LW absorption that does not lead to \htwo\ dissociation \citep{Draine1996,Burton1990,Baczynski2015}, \htwo\ formation heating \citep{Hollenbach1979,Omukai2000}, and heating by cosmic ray ionization \citep{Glassgold2012}. Gas is cooled by collisional ionization \citep{Cen1992}, collisional excitation \citep{Cen1992}, recombination \citep{Hui1997}, dielectronic recombination \citep{Black1981}, Bremsstrahlung cooling \citep{Osterbrock2006}, Compton cooling \citep{Haiman1996}, metal cooling by \textsc{cloudy} \citep{Ferland1998} above $10^4$ K and fine structure cooling \citep{Rosen1995} below $10^4$ K, and \htwo\ cooling \citep{Halle2013,Hollenbach1979}.

Photons are created through stellar injection based on stellar energy distribution (SED) tables for stellar population models.. Each star particle in \ramses\ represents an entire population of stars. The quantity of photons injected into each photon group is determined by the mass, age, and metallicity of the star particle, and is calculated from the \citet{Bruzual2003} SED tables (see \citealt{Rosdahl2013}). Photons are absorbed by dust, the photoelectric effect, and gas via either photodissociation or ionization. 

Self-shielding of \htwo\ is vital to its formation. \htwo\ dissociation occurs by line absorption in the LW band, where firstly only 10 per cent of absorptions lead to dissociation \citep{Stecher1967}, and secondly interference between absorption bands of differing strengths leads to decreasing \htwo\ destruction with increasing column density \citep{Draine1996}. Previous galaxy codes with \htwo\ have modelled this effect by decreasing \htwo\ destruction and using an approximation to convert volume density in the code to a column density \citep{Wolcott-Green2011}, with the exception of \citet{Baczynski2015} who use a ray tracing code. What we do instead is to enhance LW destruction at each time-step but not \htwo\ dissociation, and thereby do not require a conversion to column density because the LW photons will travel through the column of cells with each time-step. We calibrate the factor of LW destruction enhancement with one-dimensional simulations of radiation dissociating an \htwo\ slab (see \citealt{Nickerson2018}). We compare our transition between \htwo\ and \hi\ with its position as predicted by the analytical model of \citet{Bialy2017}, and find that enhancing the LW destruction by a factor of 400 gives good results for a range of incident fluxes, slab densities, and metallicities \citep{Nickerson2018}.

\subsection{Star formation and feedback}
\label{sec:sfrhist}

We use the star formation model described in \citet{Rasera2006} that follows a \citet{Schmidt1959} law based on the total gas density. Gas above a critical density and below $3\times10^4$ K turns into star particles following:
\begin{equation}
\dot{\rho}_*=\begin{cases}
\epsilon_{\mathrm{ff}}\rho/t_{\mathrm{ff}}&n > n_*,\\
0& n \leq n_*,
\end{cases}
\end{equation}
where $\dot{\rho}_*$ is the rate of gas conversion into stars in units of mass per volume per time, $\epsilon_{\mathrm{ff}}=0.02$ is our choice of local star formation efficiency, $\rho$ is the gas density, $t_{\mathrm{ff}}=\sqrt{3\pi/(32G\rho)}$ is the gas free-fall time,  $n_{*}$ is the star formation critical density, and $G$ is the gravitational constant. The critical density for star formation ($n_{*}$) and resulting mass of each new star particle ($m_{*}$) depend on resolution (see Table \ref{tbl:galparams}) and the number of stellar particles formed in each cell in each time-step is randomly drawn from a Poisson distribution of the \citet{Schmidt1959} law.

We use a delayed cooling model for stellar feedback as developed by \citet{Teyssier2013}. At the age of 10 Myr, each star particle releases the energy, $E_{\text{SNII}}$,
\begin{equation}
E_{\text{SNII}}=10^{51} \frac{\eta_{\mathrm{SN}}m_*}{M_{\text{SNII}}} \text{erg},
\label{eqn:snerg}
\end{equation}
where $\eta_{SN}=0.3$ is the fraction of stellar mass released by supernovae, and $M_{\text{SNII}}=10$ \msun\ is the Type II supernova mass, close to \citet{Chabrier2003}. Our supernovae do not release metals. Delayed cooling is intended to offset the effect of numerical overcooling, whereby non-thermal energy is allowed to cool over a 20 Myr time-scale.

Our homogeneous ultra violet (UV) background \citep{Faucher-Giguere2009} is dampened above densities of $10^{-2}$\cci\ to mimic gas self-shielding.

\section{Galactic Analysis}
\label{sec:gan}

In this section, we focus on analyzing three galaxies that differ by resolution: GLow, GMed, and GHigh. We present their SFR histories, morphologies, radial profiles, KS diagrams, phase diagrams, and the density and fraction distributions of \htwo.

\subsection{SFR History}
\label{ssec:sfrhist}

\begin{figure}
\includegraphics[width=0.8\columnwidth]{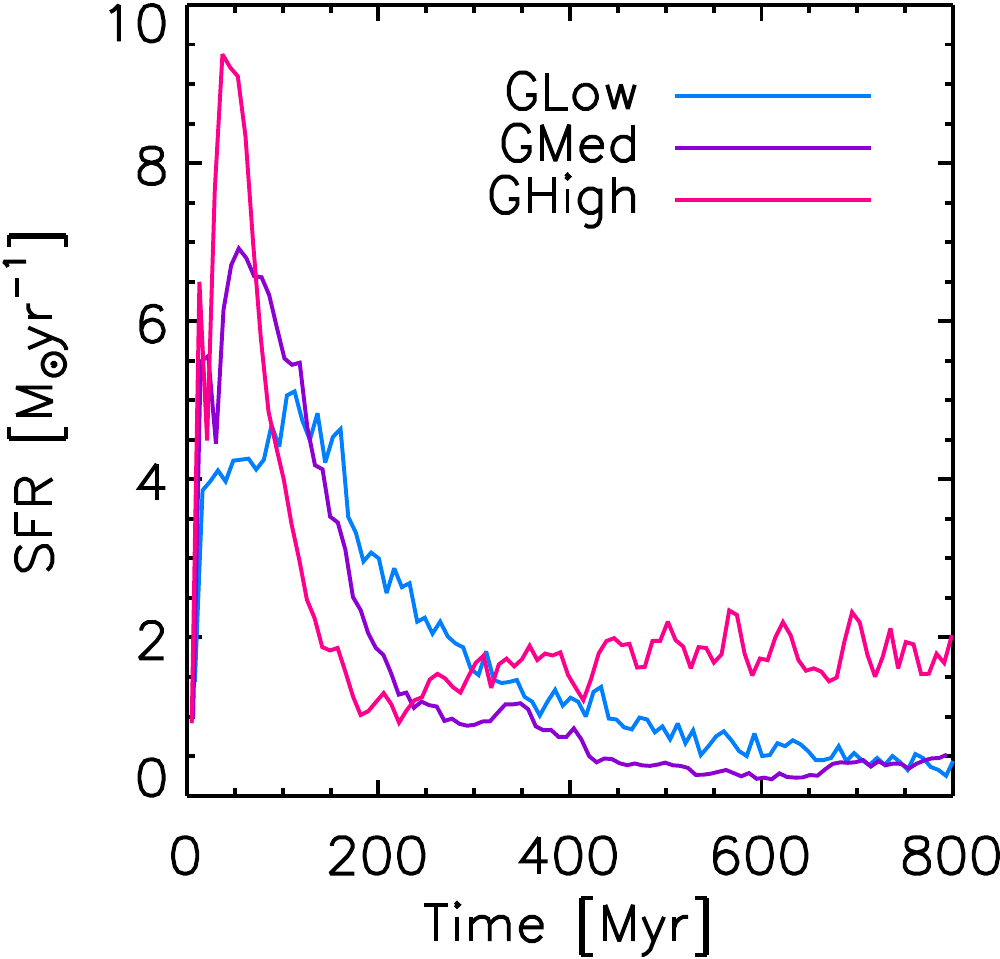}
\caption{The star formation history of GLow (blue), GMed (purple), and GHigh (pink).}
\label{fig:sfhist}
\end{figure}

We first show the discs' evolution over time. Fig. \ref{fig:sfhist} gives the star formation history of GLow, GMed, and GHigh. The discs begin with a uniform exponential structure, and quickly evolve substructure. Star formation in all three simulations is initially high for 100-200 Myr, and then calms down as the disc settles into its semi-steady state. GLow's SFR slowly decays over the course of 500 Myr until it reaches a semi-steady state at about 800 Myr. GMed, on the other hand, settles much more quickly at about 500 Myr. GHigh reaches a semi-steady state much faster following its initially high SFR at 300 Myr, and settles into a much higher SFR as compared to GMed and GLow, marked by bursts of star formation. 

We run GMed and GLow to 2 Gyr beyond what is shown here, in order ensure that 800 Myr is indeed a representative time of their equilibrium state. GHigh is too computationally expensive to run longer, but it has been stable for several hundreds of Myr. Accordingly, we choose 800 Myr as the time at which to analyze the galaxies. 

\subsection{Morphology}
\label{ssec:morph}

\begin{figure*}
\includegraphics[width=0.90\textwidth]{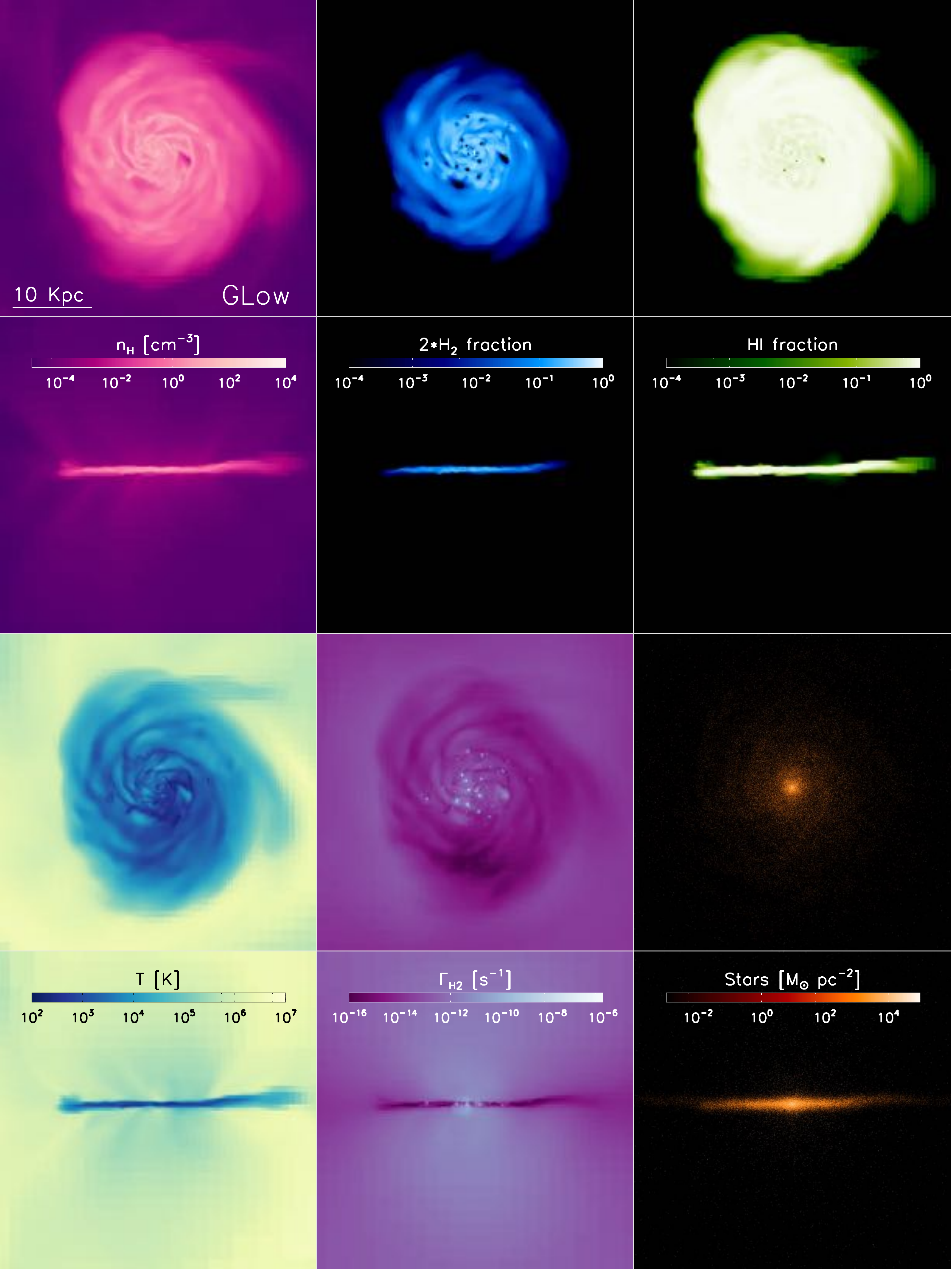}
\caption{Face-on maps of GLow mass weighted average over the disc height, and side-on maps over the disc width. Top row, left to right: face-on total gas density (\cci), \htwo\ fraction, and \hi\ fraction. Second from top row: same as top row, side-on. Second from bottom row, left to right: face-on gas temperature (K), \htwo\ photodissociation plus photoionization rate over all groups (\si), and stellar density of all stars (\msun/pc$^2$). Bottom row: same as second from bottom row, side-on. The Habing value, $G_0$, in units of \si\ for the photodestruction rate is about $6\times10^{-11}$\si.}
\label{fig:glow}
\end{figure*}

\begin{figure*}
\includegraphics[width=0.90\textwidth]{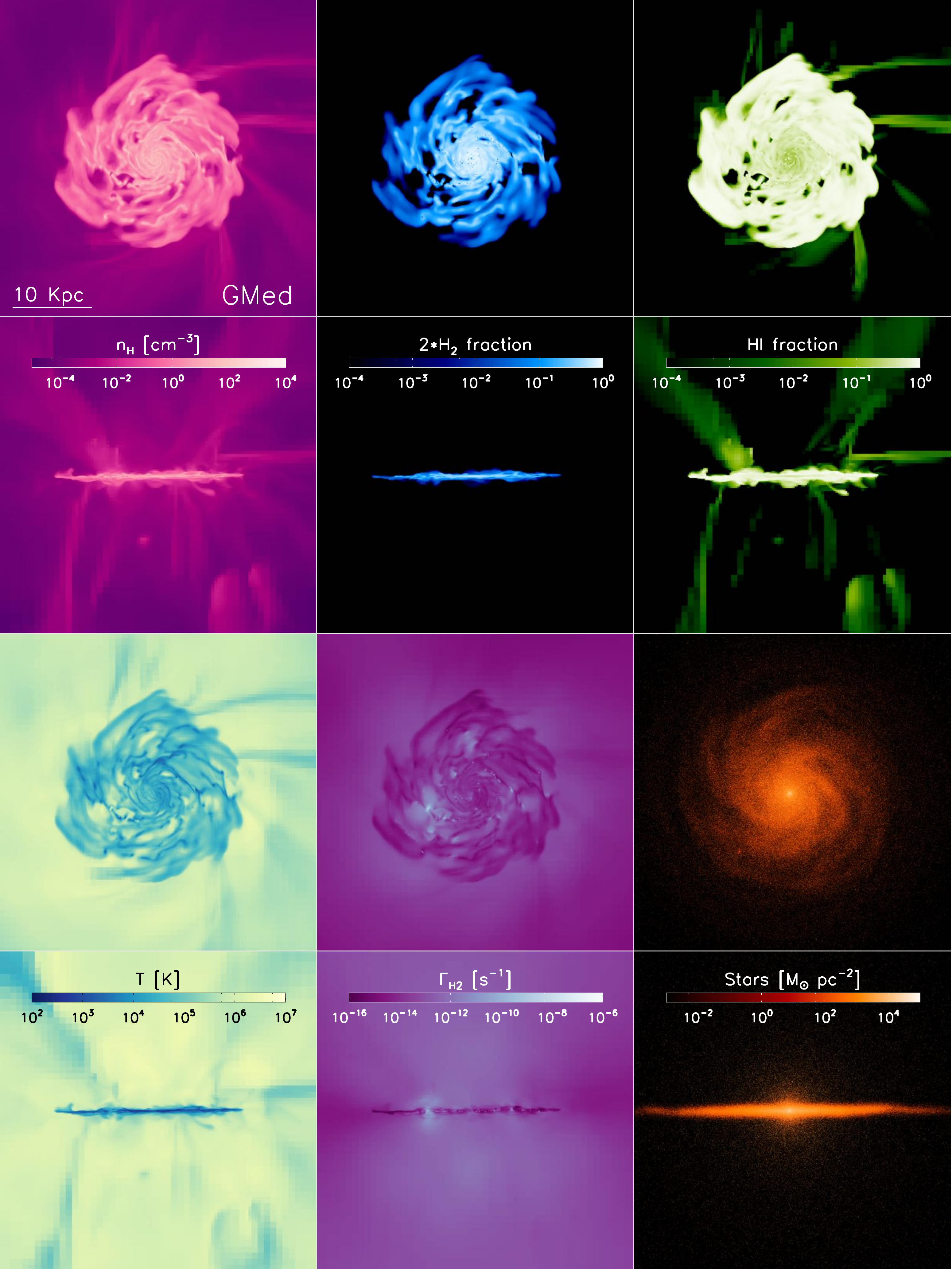}
\caption{The same as in Fig. \ref{fig:glow} for GMed.}
\label{fig:gmed}
\end{figure*}

\begin{figure*}
\includegraphics[width=0.90\textwidth]{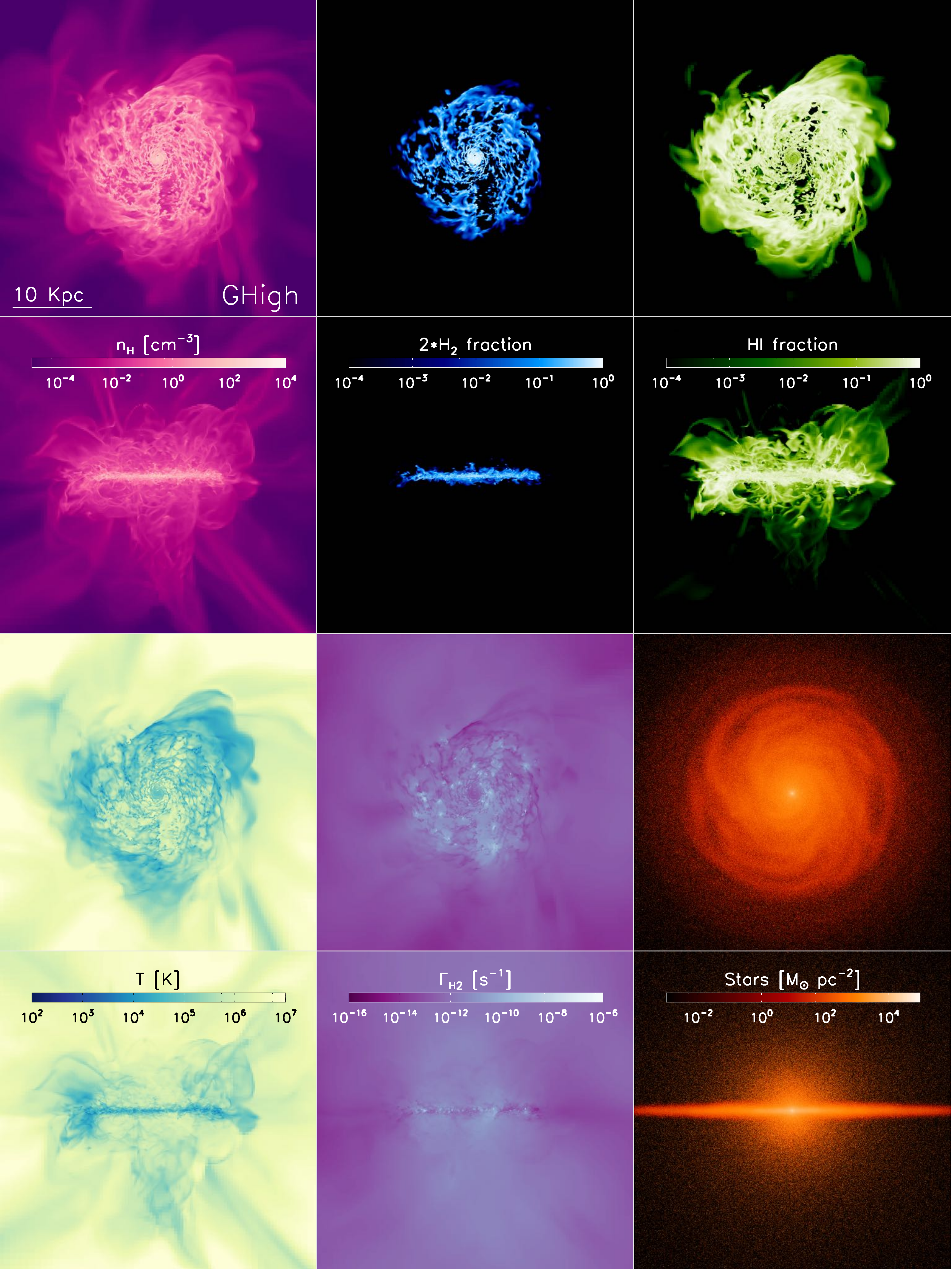}
\caption{The same as in Fig. \ref{fig:glow} for GHigh.}
\label{fig:ghigh}
\end{figure*}

In Figs. \ref{fig:glow}, \ref{fig:gmed}, and \ref{fig:ghigh} we provide maps for GLow, GMed, and GHigh face-on and side-on of the total gas density, \htwo\ fraction, \hi\ fraction, gas temperature, \htwo\ total photodissociation and ionization rate, and stellar density. We do not include the \hi\ photoionization rate because it follows the \htwo\ rate very closely, both sourced from young stars and absorbed by gas. 

We define the maximum \htwo\ abundance fraction $\xhtwo=0.5$ because \htwo\ is diatomic. In in this work we display $2*\xhtwo$ so that its range from 0 to 1 is visually comparable to $\xhi$ and $\xhii$.

The most obvious difference between the different resolution simulations firstly comes in the structure of the gas density. The gas is much more diffuse and the spiral arms fewer in GLow, while the complexity of the gas increases in GMed and GHigh. The high-density region, 10 \cci\ and higher, is confined to a smaller central core in GLow, and this is larger in GMed and GHigh. GHigh has more spiral arms, which are thinner, and filamentary structures that are present in neither GMed nor GLow. The gas in GHigh has many more clumps, a subject we will return to in Section \ref{sec:mc}. The regions of high-density gas are also sharpest in GHigh, with a more distinct envelope of $\approx 10^{-3}$ \cci\ gas that ends sharply at $\approx 10^{-5}$ \cci. In the side-on maps, we can see material ejected above and below the disc by supernovae feedback. Very little is ejected in GLow, while streams of gas emit from GMed. GHigh has the most elaborate structure in its ejected gas, showing an additional fountain of high-density gas that soon falls back onto the disc. 

The resolution differences in gas densities translates into differences in the \htwo\ and \hi\ face-on maps. \htwo\ traces the densest regions in the galaxy, and accordingly increases in structural complexity with resolution. While GMed's \htwo\ map is a simple progression of GLow, being denser and having more arms, GHigh brings new features to the \htwo\ map. The outer \htwo\ regions are much thicker and diffuse, while the middle region has more numerous \htwo-free pockets and thinner \htwo\ clouds as compared with GMed. The innermost disc returns to being continuously high in \htwo\ with a delicate spiral structure. \hi\ is more spread out and evenly distributed than \htwo\ and is present in the inter-arm regions. It follows the same resolution effects as \htwo, gaining structure with higher resolution, but gaining holes mid-disc in GHigh. \hi\ traces the ejected gas and the galactic fountain as seen in the gas density maps, while \htwo\ is confined to the disc. This same effect is seen in the ISM box simulations of \citet{Girichidis2016}. 

The temperature map of GLow is much colder than that of GMed or GHigh. GMed features cold clumps and arms with hotter inter-arm regions, while GHigh has more hot patches and a distinct cold, outer envelope. High temperature regions envelop the young stars. The side-on view reveals that most of the cold gas remains in the disc in GLow and GMed. GHigh has a much richer side-on view, the disc showing a cold fountain of gas, also seen in \hi, expelled from the disc. 

The photodestruction maps follow the dissociating and ionizing radiation of the youngest stars and any supernovae, and accordingly tend to vary the most depending on which time-step we use. Generally, GHigh does have more stars and radiation spots of varying sizes spread throughout most of the disc. GMed has fainter radiation throughout the disc, while GLow has a brighter central concentration of stars as compared to GMed, but very little in the rest of the disc. The dark swathes in these maps correspond to the highs in the \htwo\ map, where it successfully shields against the photons. Side-on, this radiation leaks out above and below the disc, but very little penetrates the plane of the disc.

The stellar density map includes all stars. The density and structural complexity of the stellar map increases greatly with resolution. Spiral structure is barely resolved in GLow with mostly centrally concentrated stars, but becomes apparent in GMed, with GHigh featuring an even grander spiral structure. Comparing the side-on view of the stars and the gas density in GMed and GHigh, the stellar disc extends out further than the high-density region of the gas disc. Further comparison with the radiation map, which traces young stars, shows that the young stars are centrally concentrated, and the outer disc tends to be older stars. GMed and GHigh develop bulges of older stars. In GLow, the old stars trace the gas disc and do not extend beyond it.


\subsection{Galactic observables}
\label{ssec:obs}

In this section we aim to provide the observable features of our galaxies. We compare our simulations to nearby galaxy measurements in the THINGS survey of \hi\ \citep{Walter2008} and the molecular surveys HERACLES \citep{Leroy2009} and BIMA SONG \citep{Helfer2003} combined in \citet{Bigiel2008}. We bin the radial profiles of our galaxies into 0.5 kpc wide cylindrical segments and take the SFR over an average of 200 Myr as recommended by \citet{Gnedin2011} for comparing simulations to these measurements.

\begin{figure}
\begin{tabular}{l}
\includegraphics[width=0.9\columnwidth]{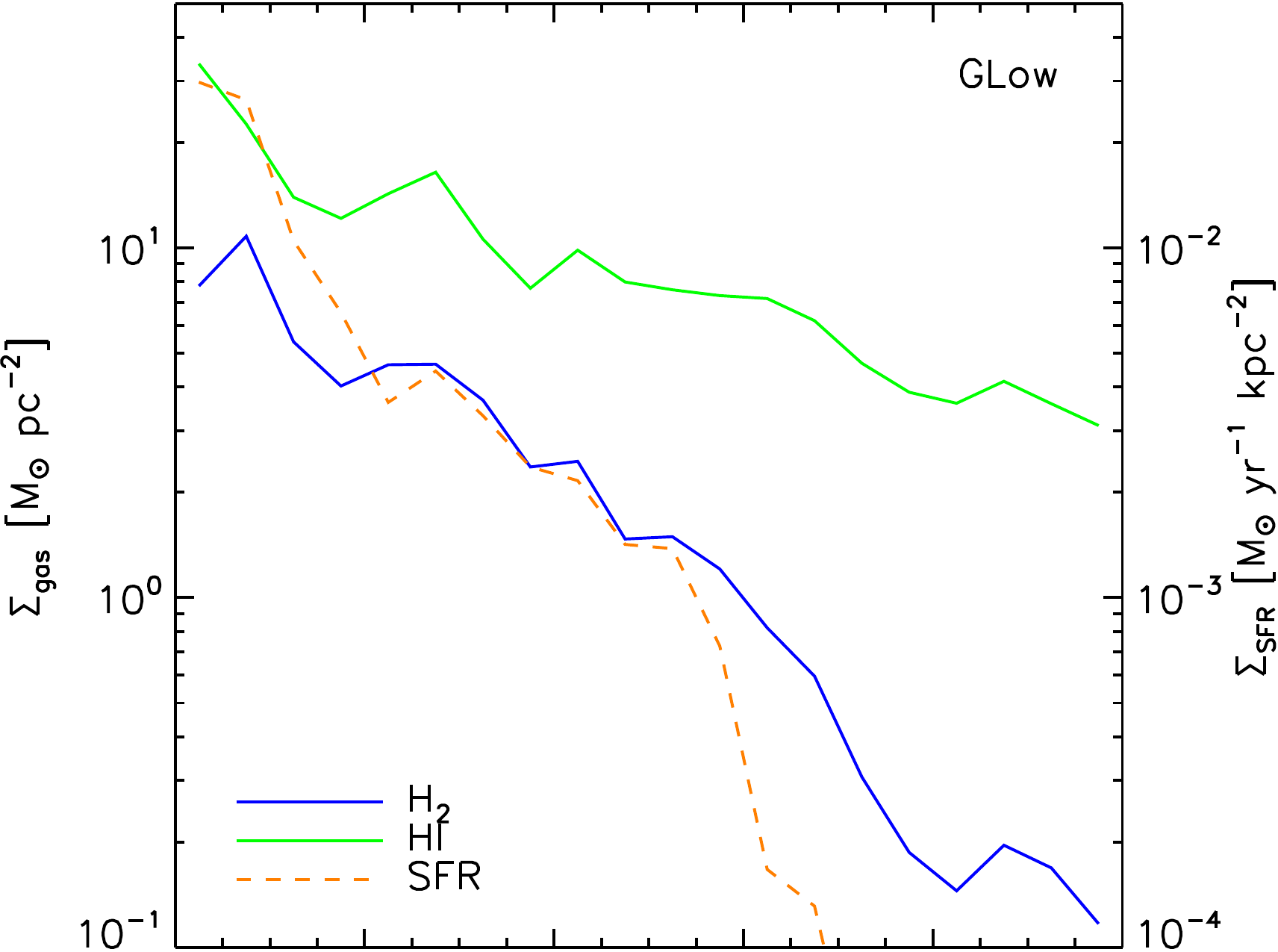} \\
\includegraphics[width=0.9\columnwidth]{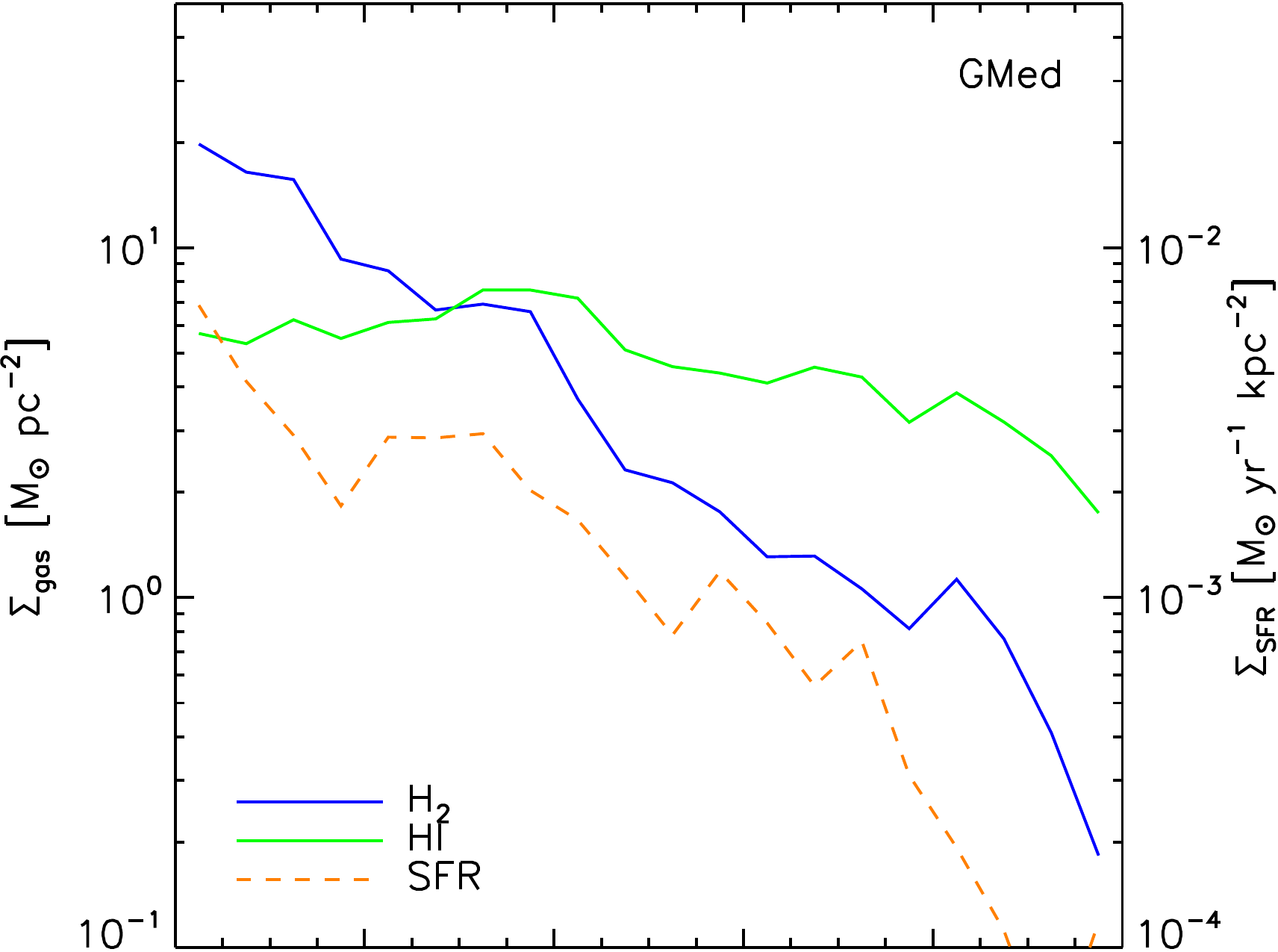} \\
\includegraphics[width=0.9\columnwidth]{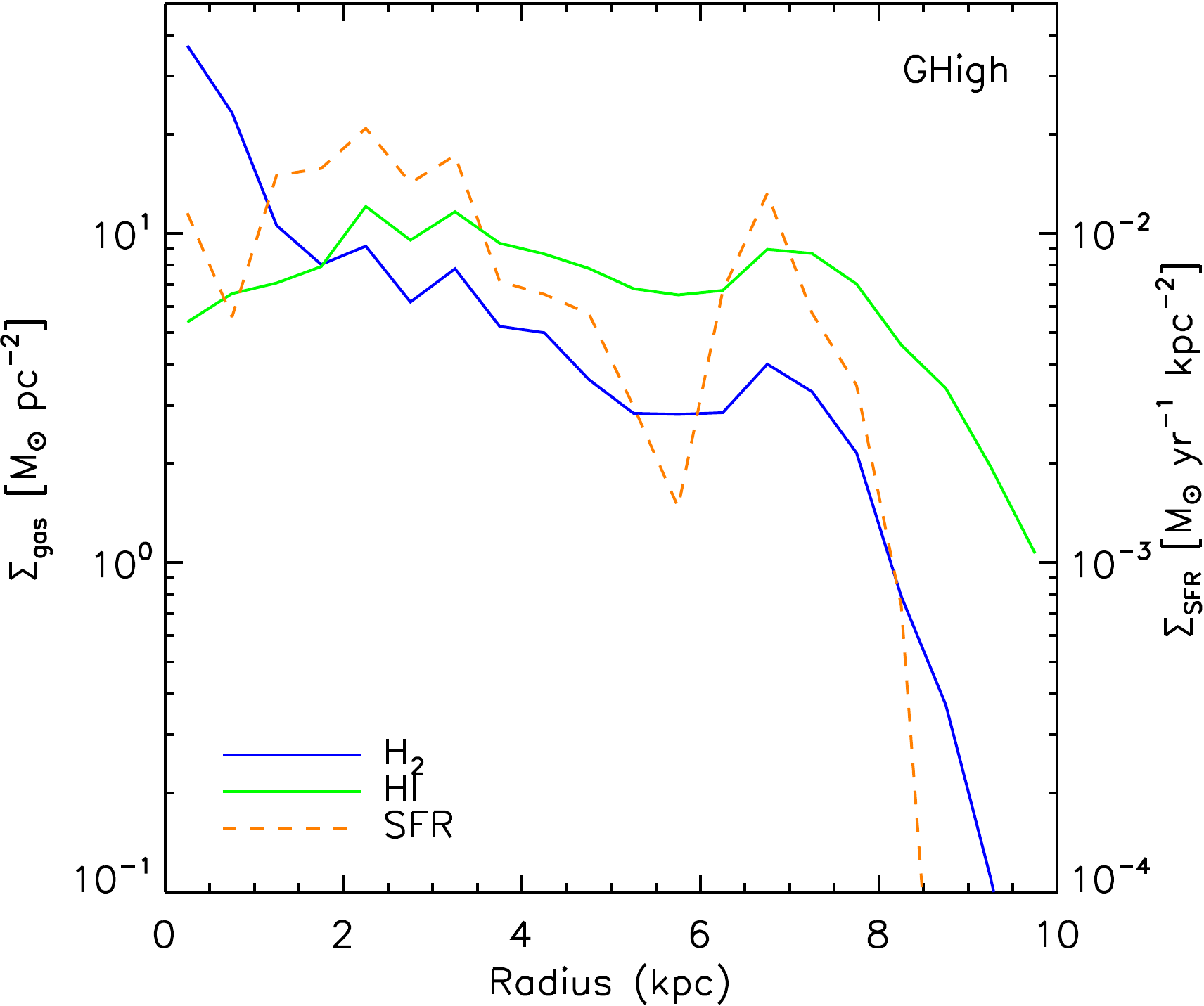}
\end{tabular}
\caption{The surface density profiles of \htwo, \hi\ (\msun pc$^{-2}$), and SFR (\msun yr$^{-1}$ kpc$^{-2}$) for GLow, GMed, and GHigh  in 0.5 kpc bins. The SFR is averaged for 200 Myr.} \label{fig:rprof}
\end{figure}

First we present the radial profiles for our three galaxies in \htwo, \hi, and SFR surface densities in Fig. \ref{fig:rprof}. Our goal is not to reproduce any one specific galaxy but to ensure that our galaxies are morphologically feasible. We compare our galaxies to the radial profiles provided in \citet{Leroy2008} and \citet{Gallagher2018}. The general trend in their observations is for \htwo\ and SFR to follow one another, peak at the galactic centre, and fall toward the outer disc, while \hi\ is roughly constant throughout the disc. In the \citet{Leroy2008} galaxy sample, \htwo\ dominates in the centre while \hi\ dominates the outer disc, while in the \citet{Gallagher2018} catalogue \htwo\ maintains dominance throughout the entire disc. 

With our galaxies in Fig. \ref{fig:rprof}, it is clear that resolution affects the profiles greatly. In GLow, the surface density of \htwo\ never overtakes \hi\ and remains low. Furthermore, the SFR peaks in the centre much higher than GMed and GHigh and drops rapidly mid-disc. This high central SFR is probably because GLow lacks a stellar bulge (Fig. \ref{fig:glow}), while the bulge is prevalent in GMed (Fig. \ref{fig:gmed}) and GHigh (Fig. \ref{fig:ghigh}). This leads to morphological quenching  in which the bulge stabilizes the gas disc against star formation \citep{Martig2009}. GMed and GHigh resemble the features of observed galaxies much more closely, especially the \citet{Leroy2008} galaxies. In both, the \htwo\ peaks in the galactic centre and falls toward the outer disc, while the \hi\ profile remains relatively flat throughout the disc. The SFR also peaks in the centre for GMed, while GHigh's SFR does not centrally peak but does have higher central \htwo\ compared to GMed. In all three of our galaxies, the trends in SFR do largely follow \htwo.

\begin{figure*}
\begin{tabular}{l|}
\includegraphics[height=0.2\paperheight]{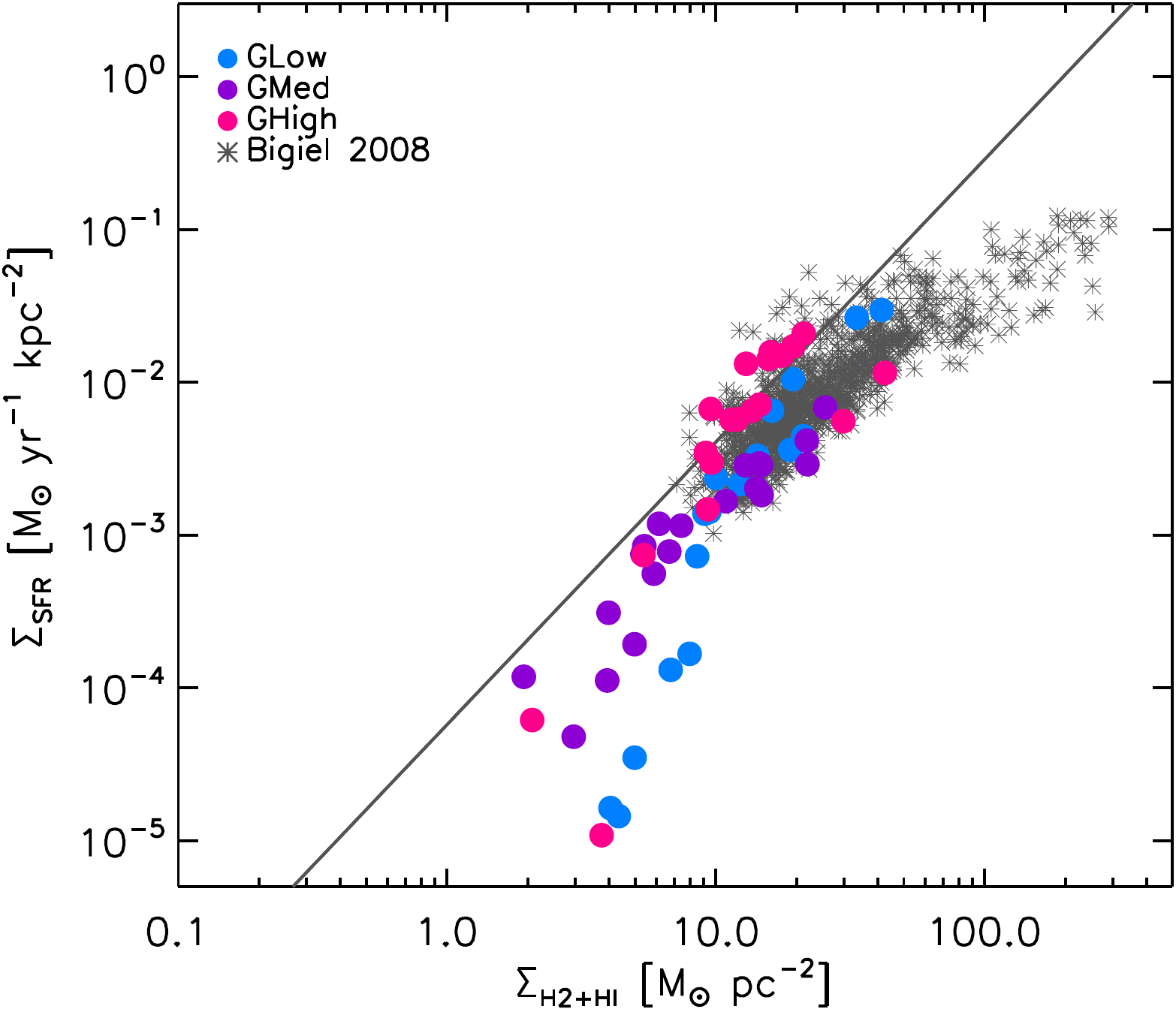} 
\includegraphics[height=0.2\paperheight]{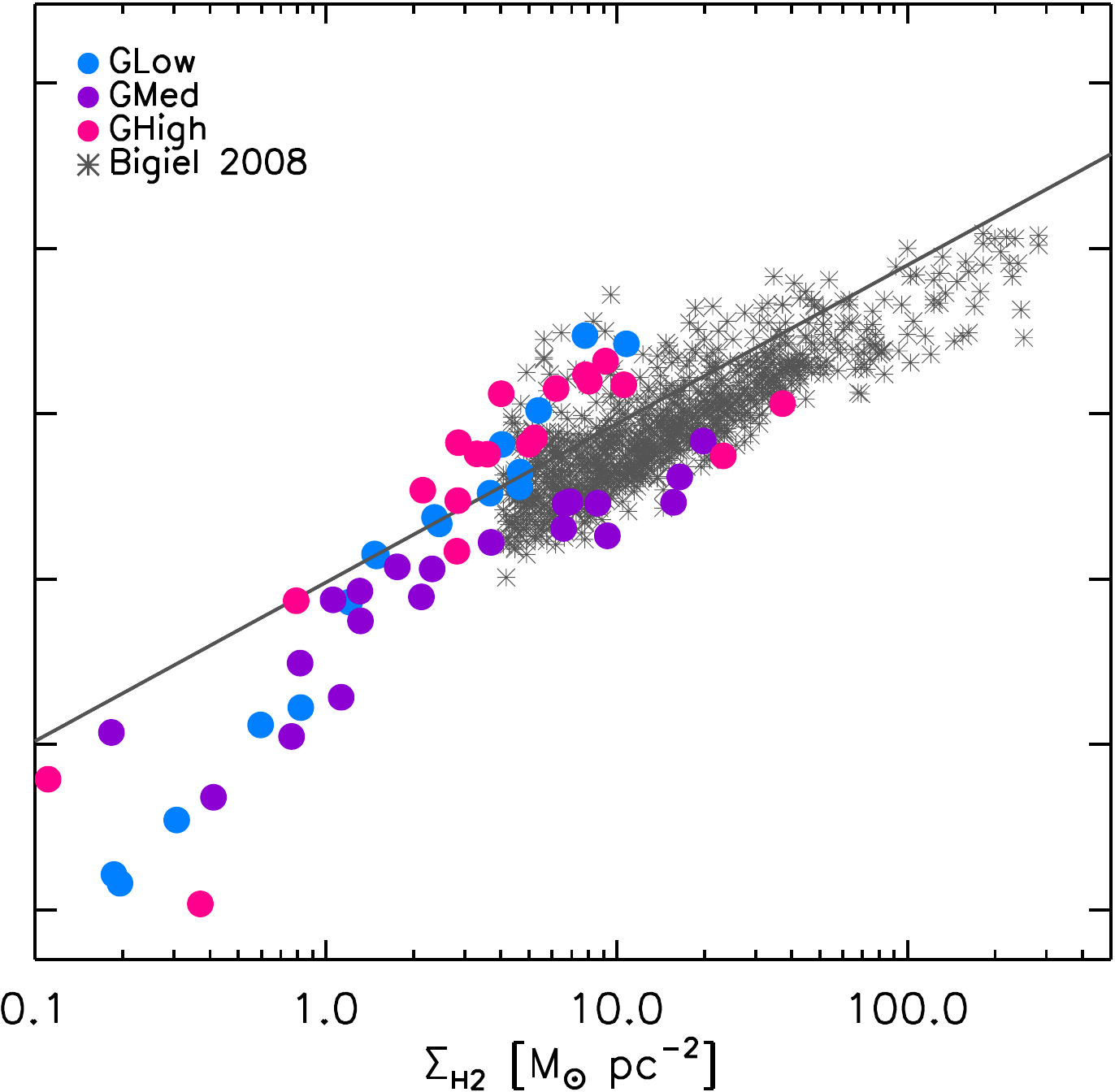} 
\end{tabular}
\caption{The Kennicutt-Schmidt relation for GLow (blue), GMed (purple), and GHigh (pink)  in 0.5 kpc bins. The SFR averaged for 200 Myr. Our simulations are compared to data from \citet{Bigiel2008} given by the grey stars, and the grey line is the power law that they derive.} 
\label{fig:ksr}
\end{figure*}

In Fig. \ref{fig:ksr} we show the KS relation \citep{Schmidt1959,Kennicutt1998} for our galaxies as compared to the data downloaded from \citet{Bigiel2008}. This snapshot at 800 Myr is typical of the relation across many snapshots for the semi-stable state of the disc.

First we give the traditional relation between the surface densities of total neutral hydrogen and SFR, and then we show the same relation between the surface densities of \htwo\ and the SFR. It is clear in both relations that GLow follows a different slope than GMed and GHigh. GHigh is a higher SFR version of GMed, sharing similar slopes. Also in both relations, our simulated galaxies fall within the margins of the observational data. Our simulations extend to much lower densities than is possible with \citet{Bigiel2008}'s instrumentation, but we cannot resolve surface densities as high  as the observations, and so observation and simulation meets in the middle. In the total neutral gas relation we reproduce not only the linear correlation at higher densities but also its breakdown at lower densities. This breakdown is not visible in the \citet{Bigiel2008} data, but is present in the later \citet{Schruba2011} data which goes to lower surface densities. However, in the \htwo\ relation each galaxy is able to maintain a roughly constant linear relation at all densities. The exception is two central points in GHigh, where our central SFR is morphologically quenched despite high $\Sigma_{\text{\htwo}}$. Our simulated galaxies reproduce the tighter correlation between the molecular gas and star formation rate as compared to neutral gas that is seen in observation.

Summing the total \htwo\ fraction of neutral gas for GLow, GMed, and GHigh we have 0.14, 0.35, and 0.33. The \htwo\ fraction of neutral gas within the solar circle (defined as 7 kpc) is about 0.25 to 0.29, taking the \htwo\ mass as estimated by \citet{Heyer2015} and \hi\ mass measured by \citet{Sofue2018}. GMed and GHigh are close to this, while GLow greatly underestimates the molecular content. Another way to slice the data is to take the ratio of \htwo\ mass to \hi\ mass, where we find for GLow, GMed, and GHigh values of 0.17, 0.55, and 0.49 respectively, which all fall into the range for disc galaxies in the COLD GASS survey by \citet{Boselli2014}, about 0.03 to 10.

\subsection{Phase diagrams}
\label{ssec:phase}

\begin{figure*}
\begin{tabular}{l||}
\includegraphics[height=0.15\paperheight]{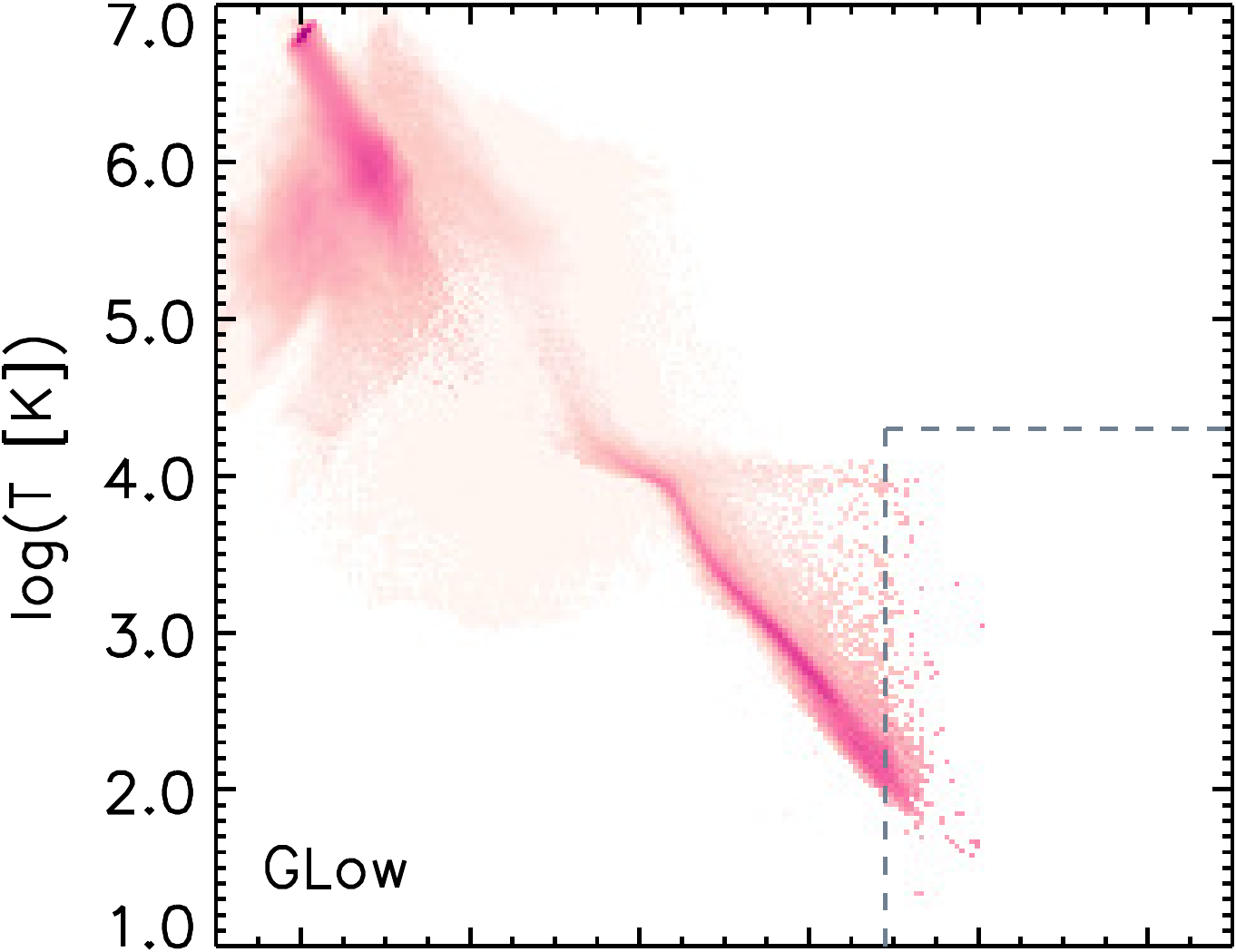} 
\includegraphics[height=0.15\paperheight]{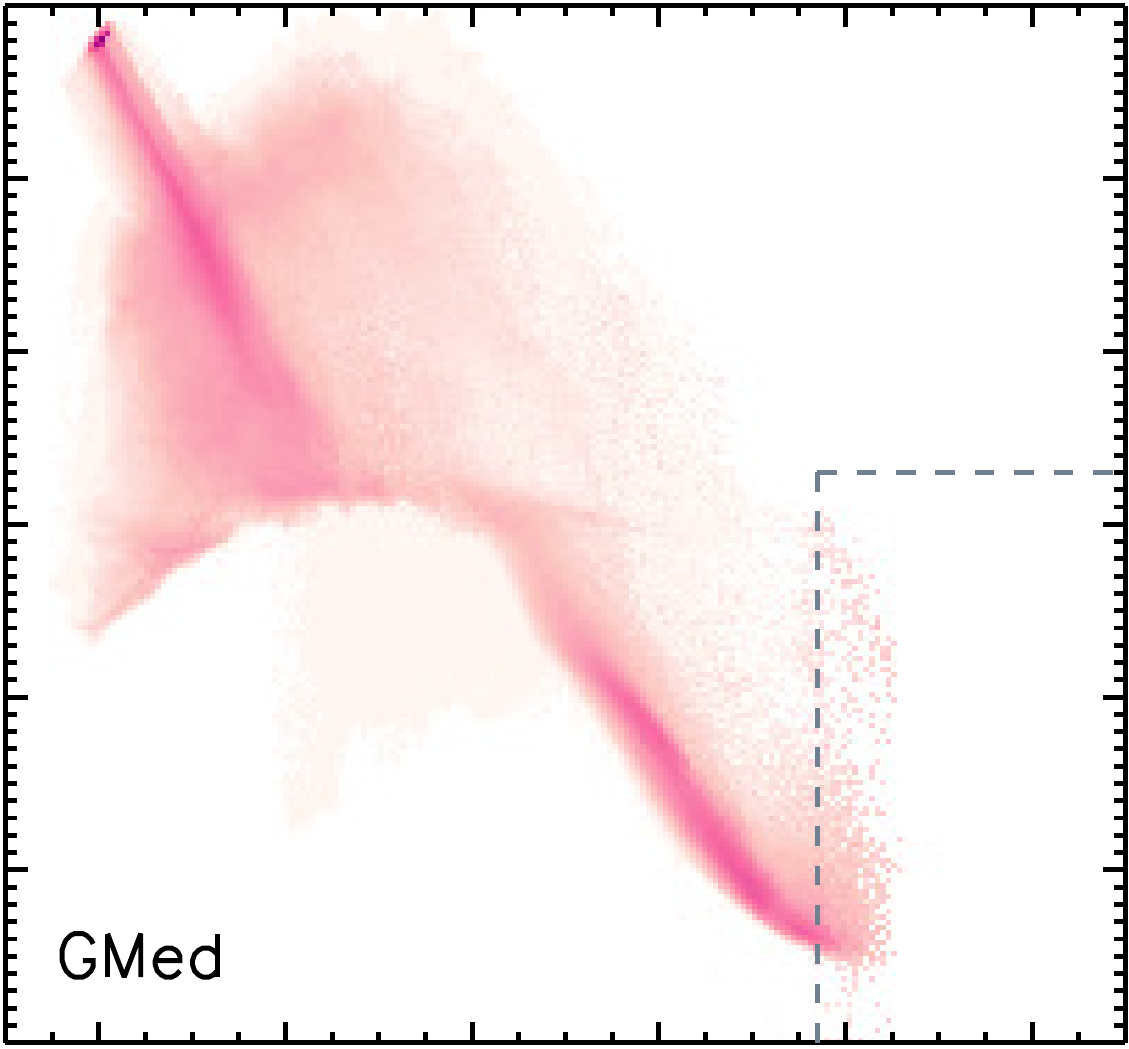} 
\includegraphics[height=0.15\paperheight]{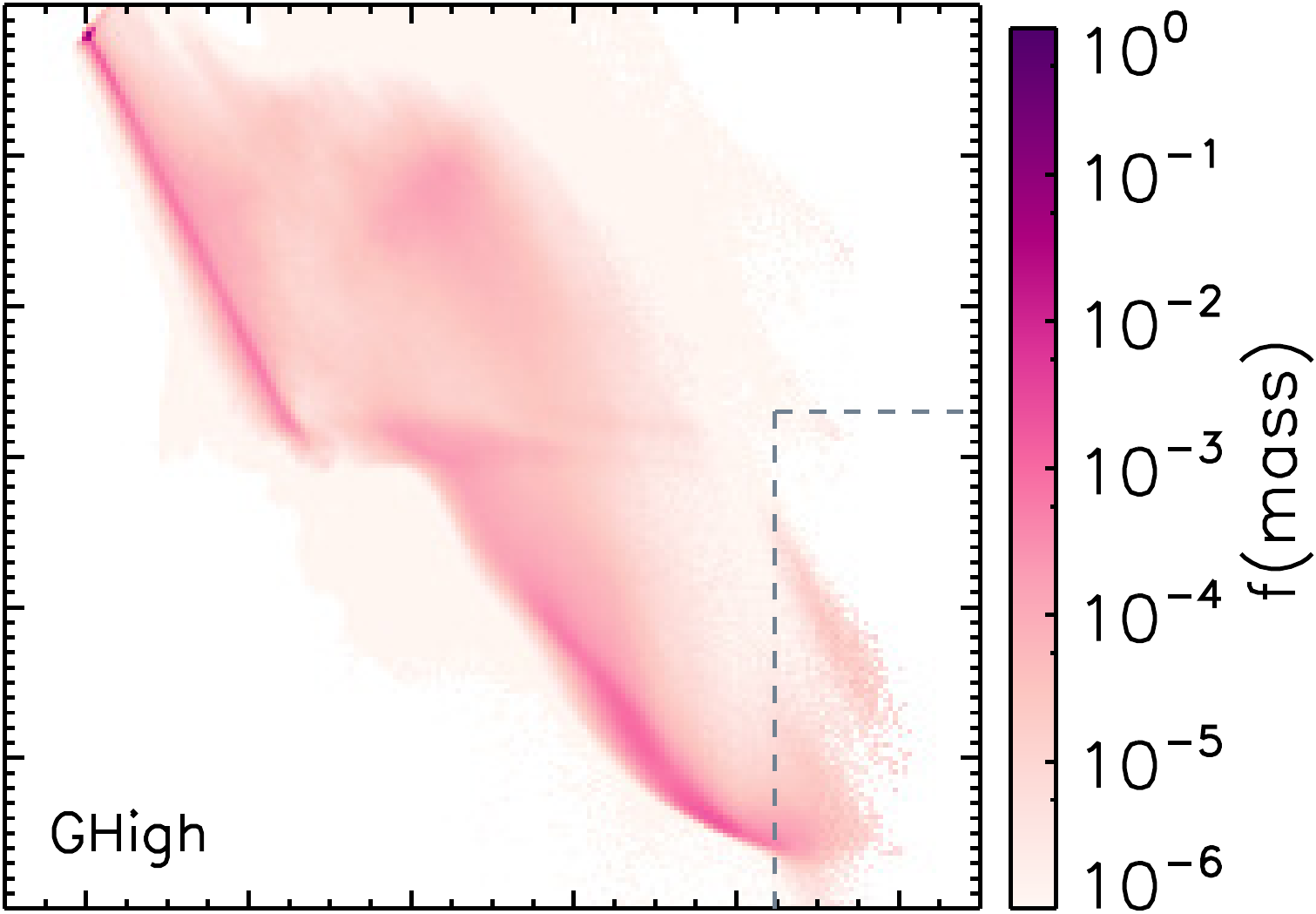} \\
\includegraphics[height=0.15\paperheight]{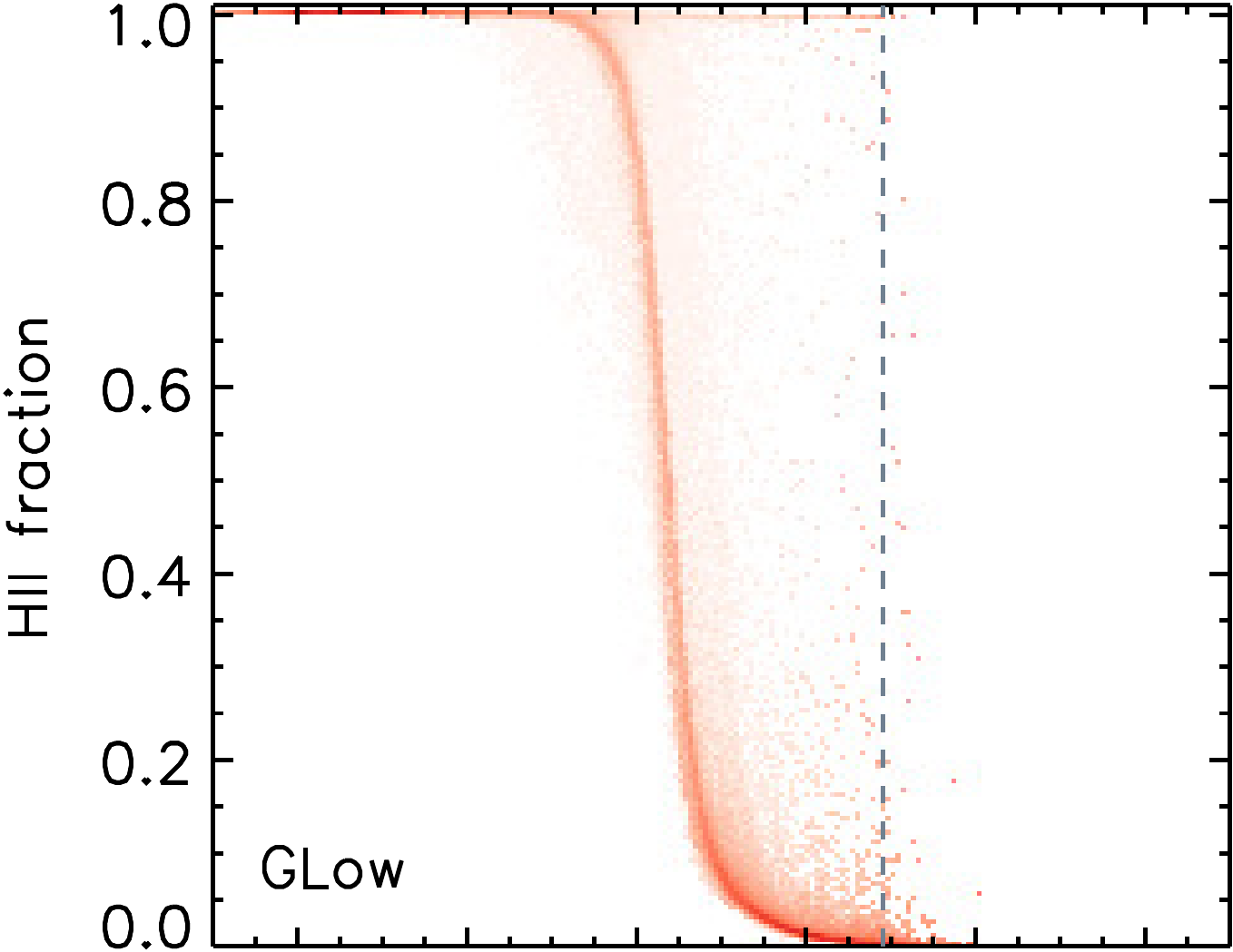} 
\includegraphics[height=0.15\paperheight]{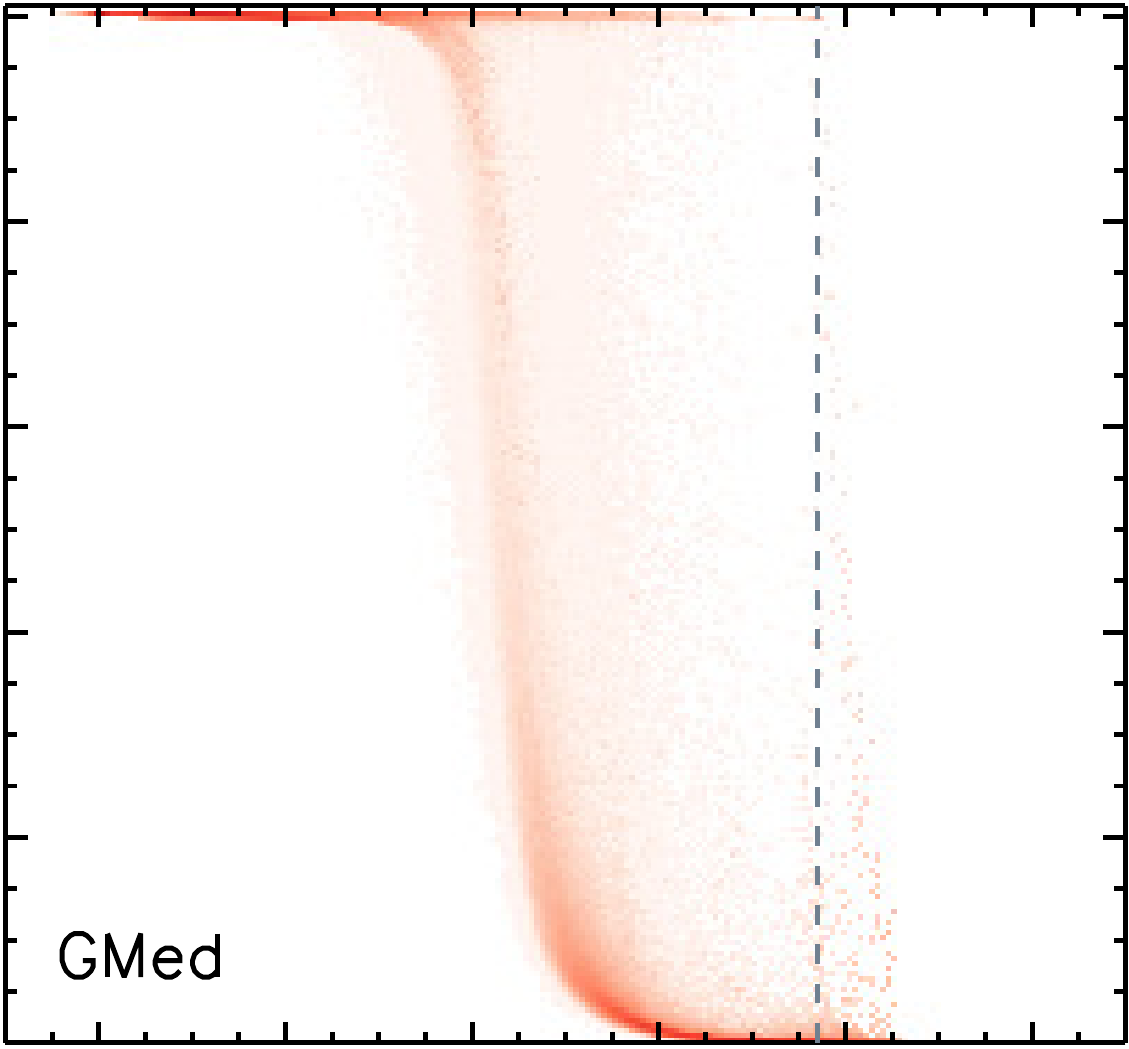} 
\includegraphics[height=0.15\paperheight]{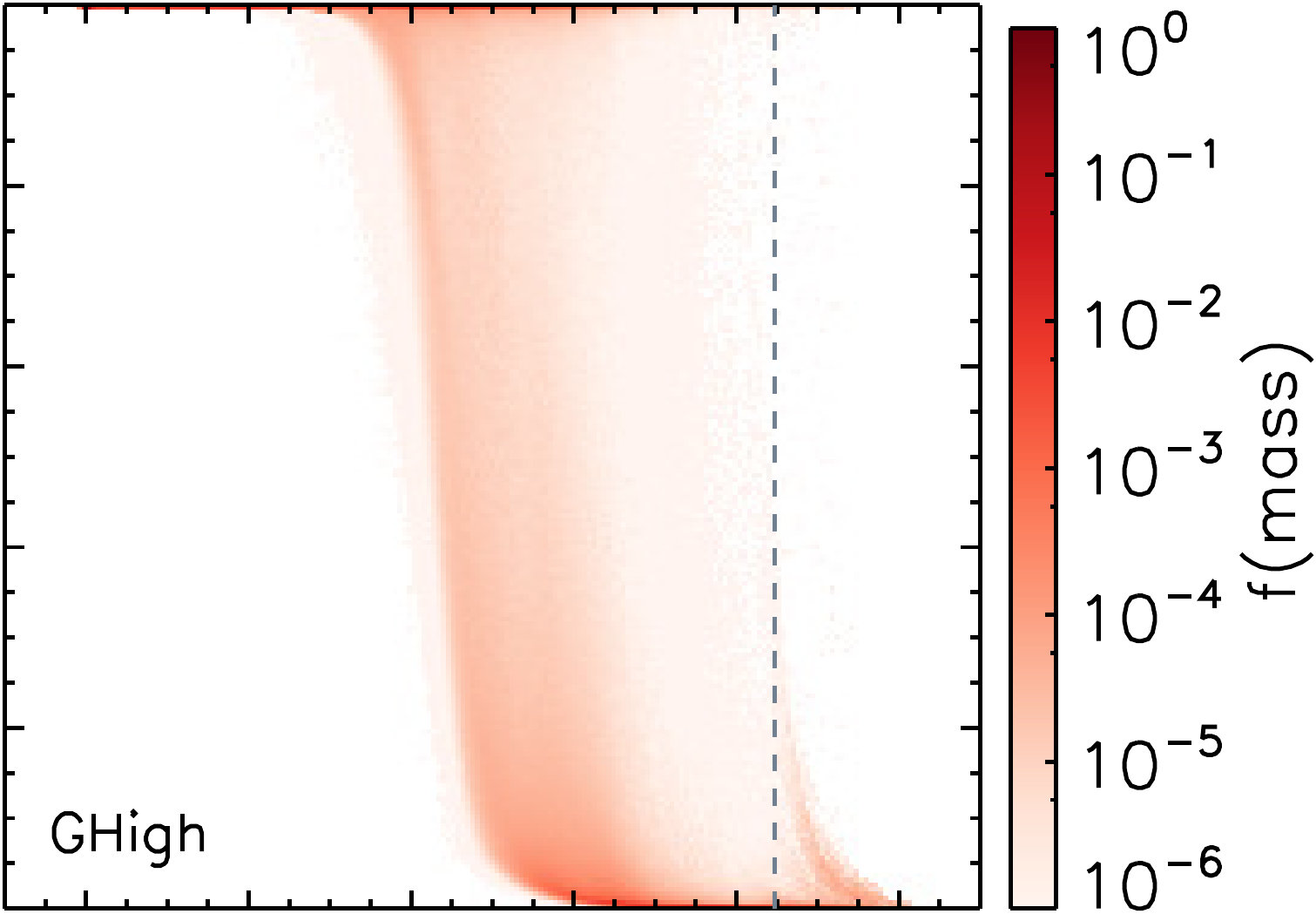} \\
\includegraphics[height=0.15\paperheight]{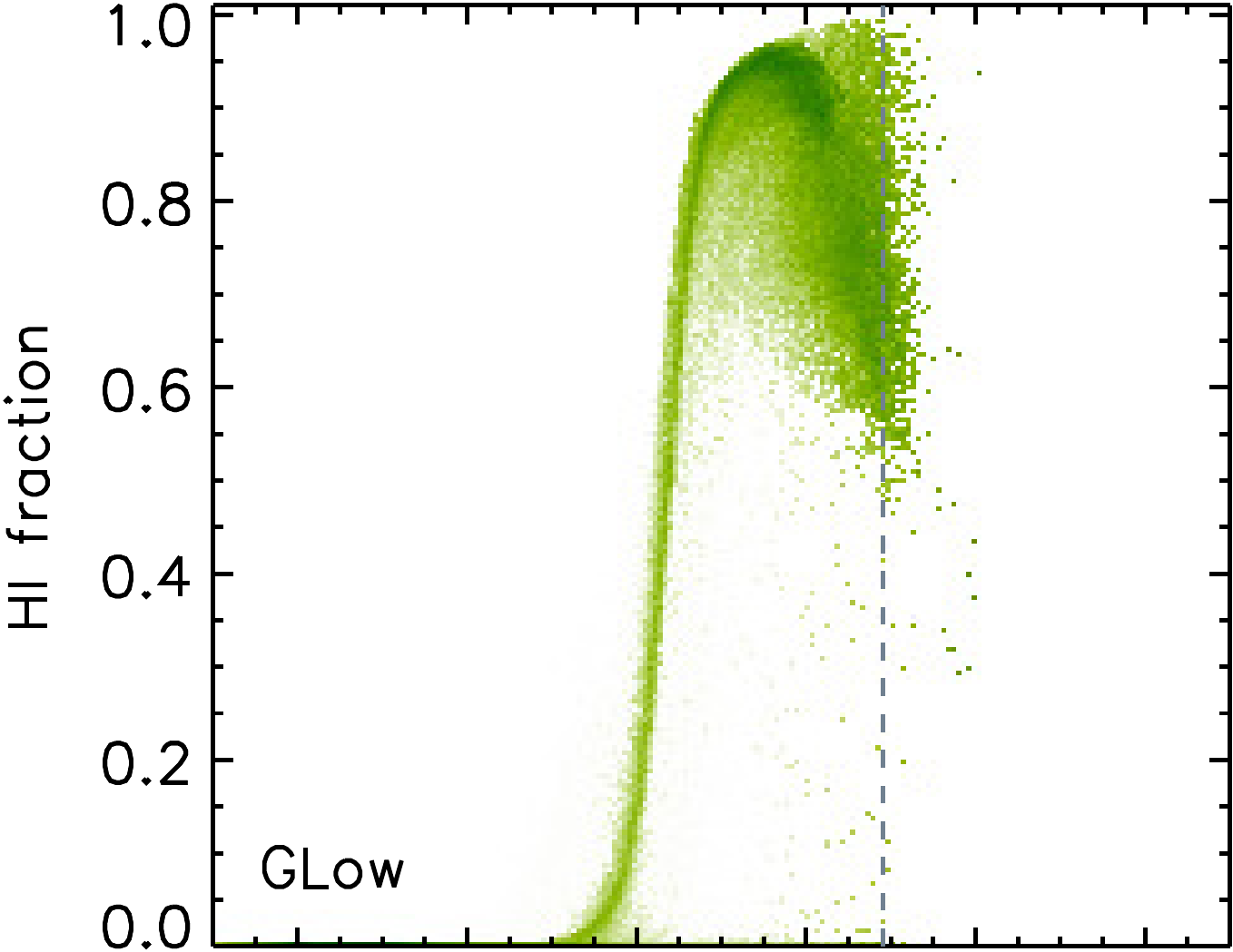} 
\includegraphics[height=0.15\paperheight]{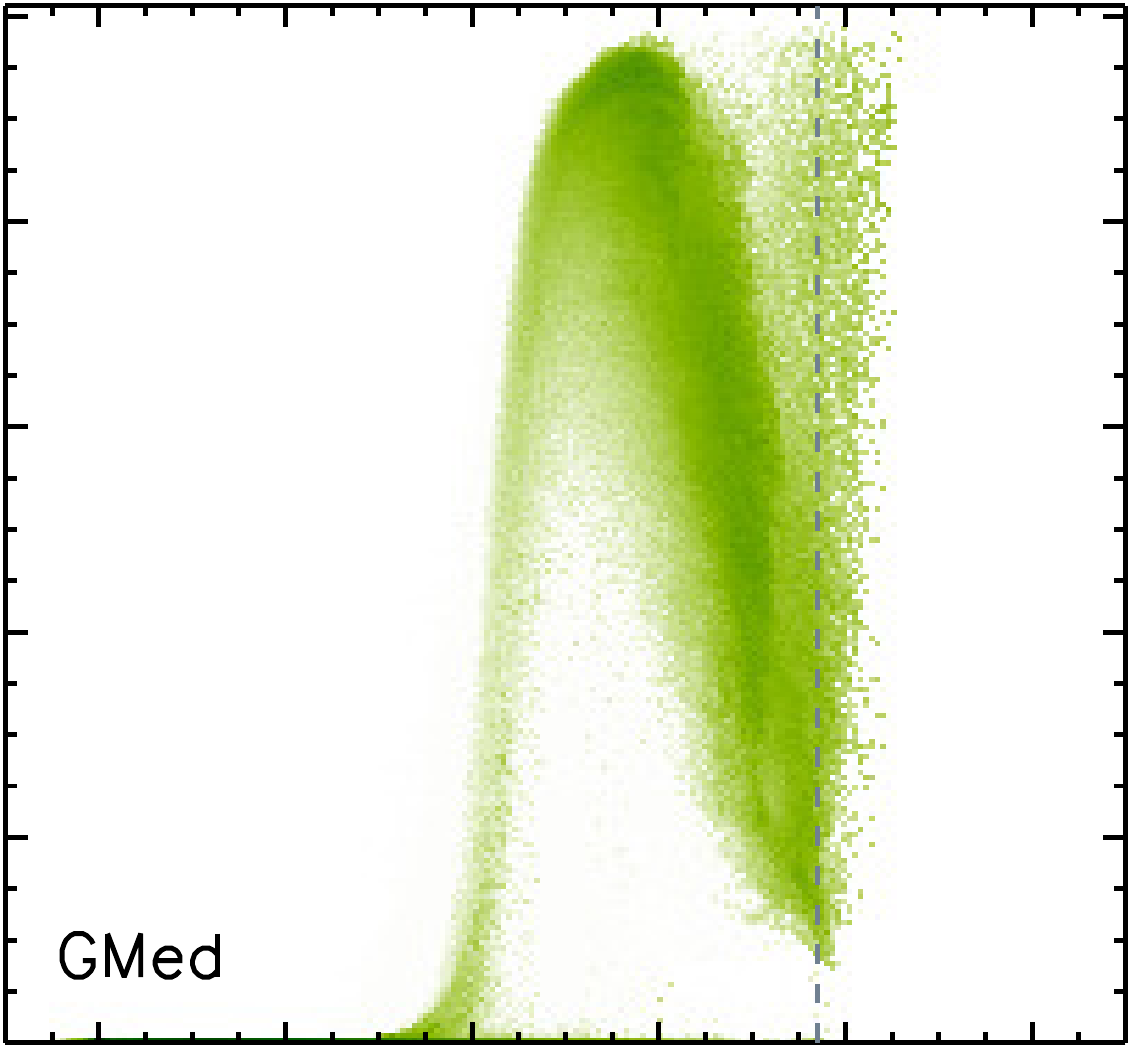} 
\includegraphics[height=0.15\paperheight]{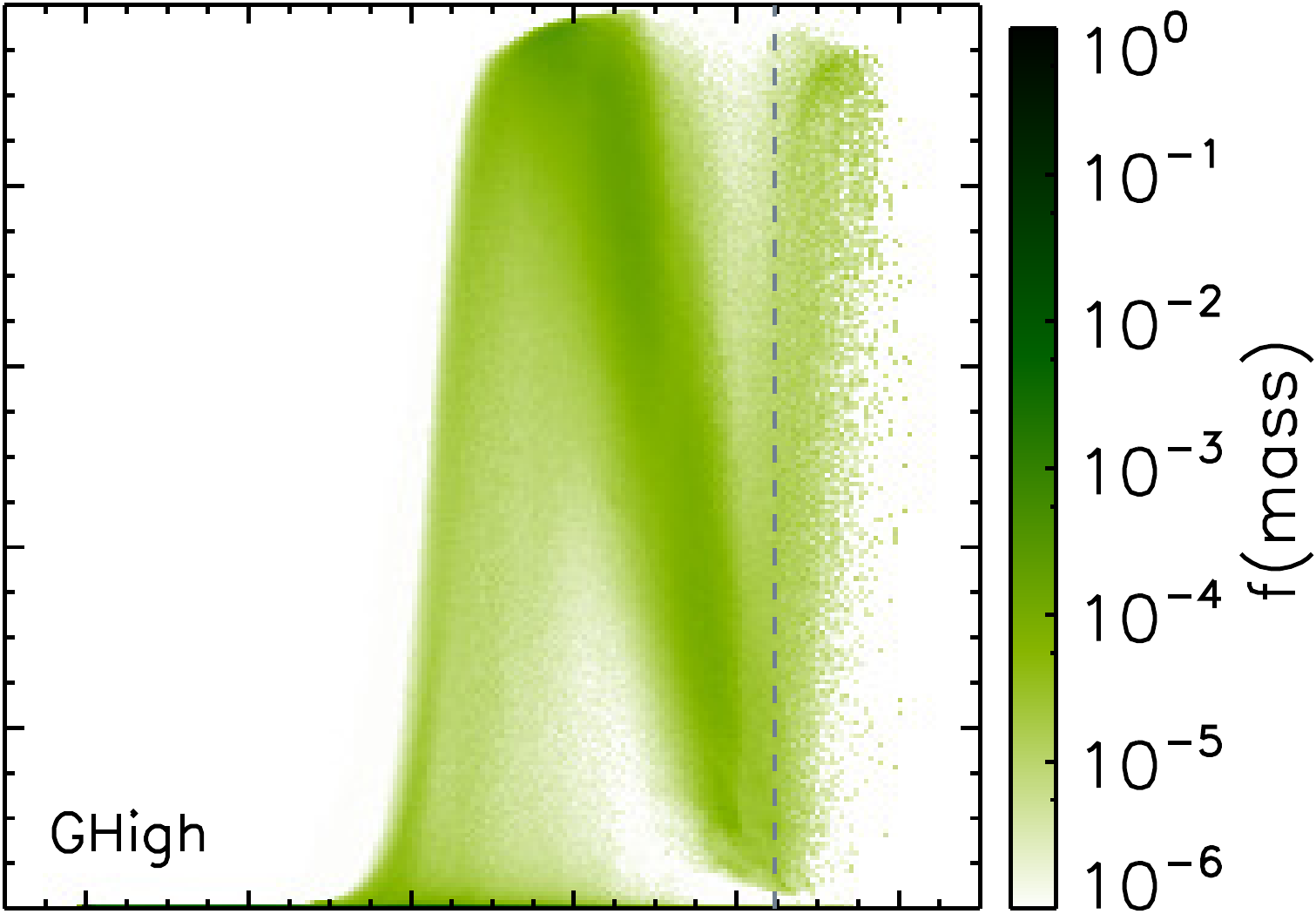} \\
\includegraphics[height=0.177\paperheight]{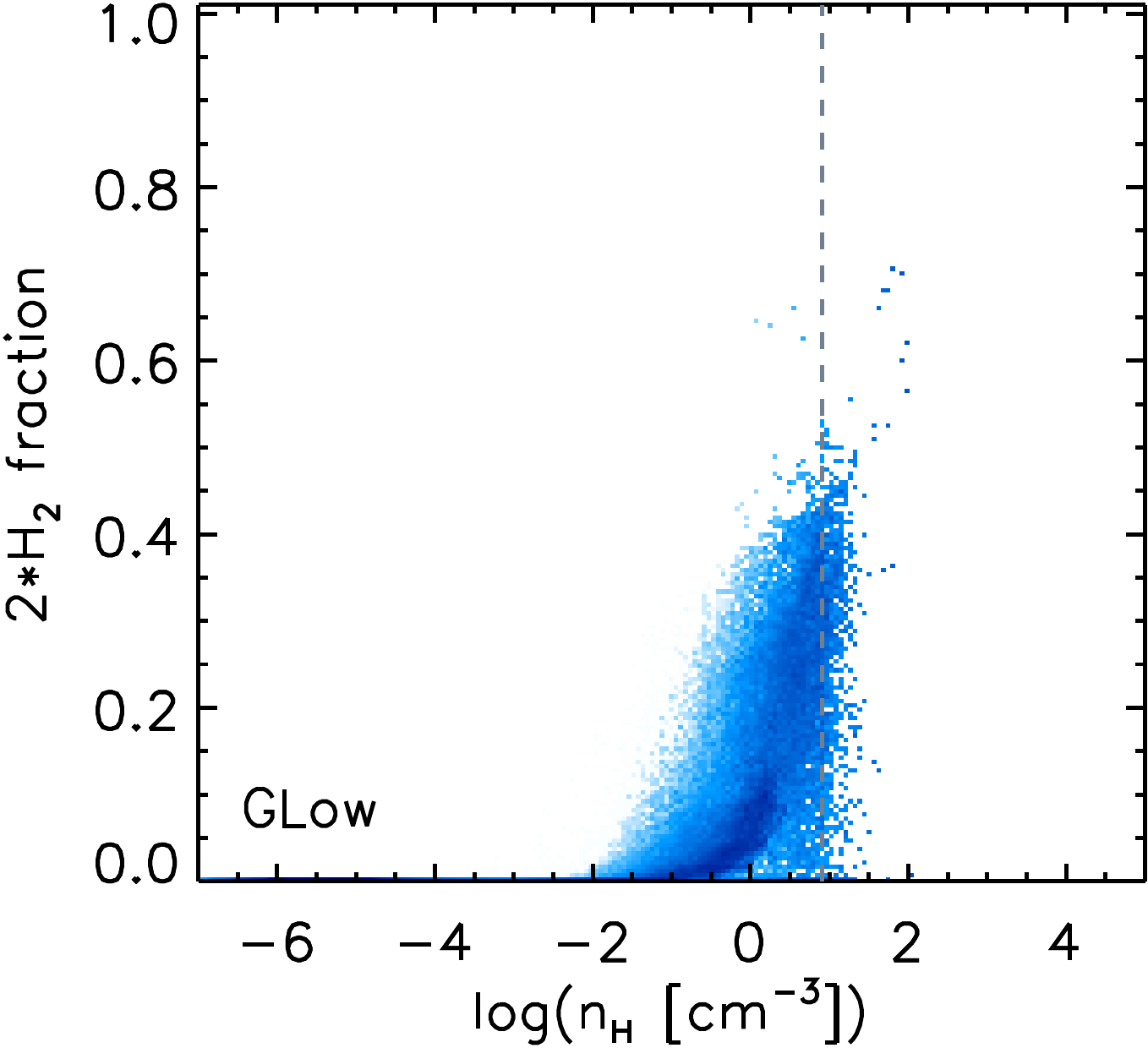} 
\includegraphics[height=0.177\paperheight]{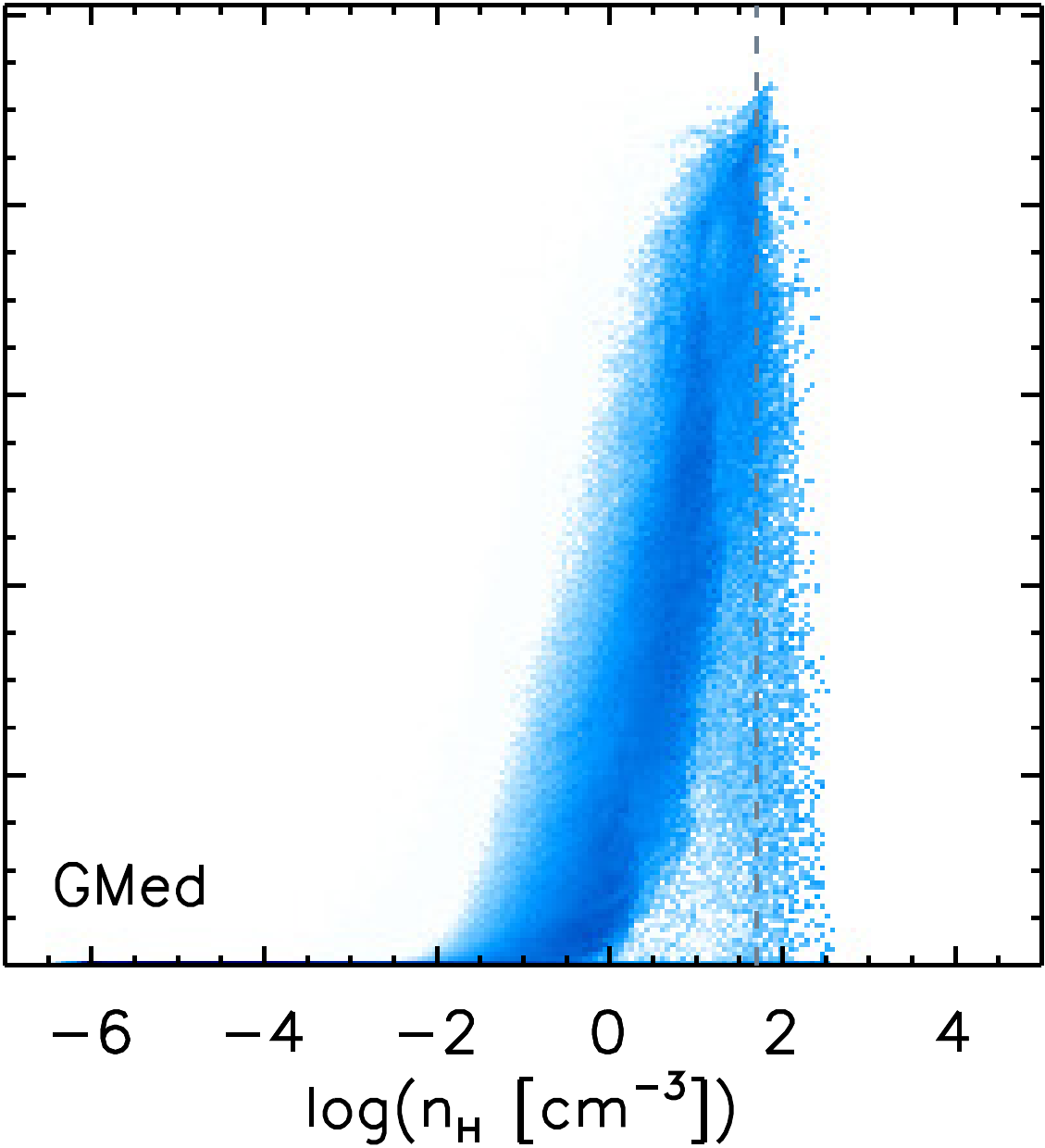} 
\includegraphics[height=0.177\paperheight]{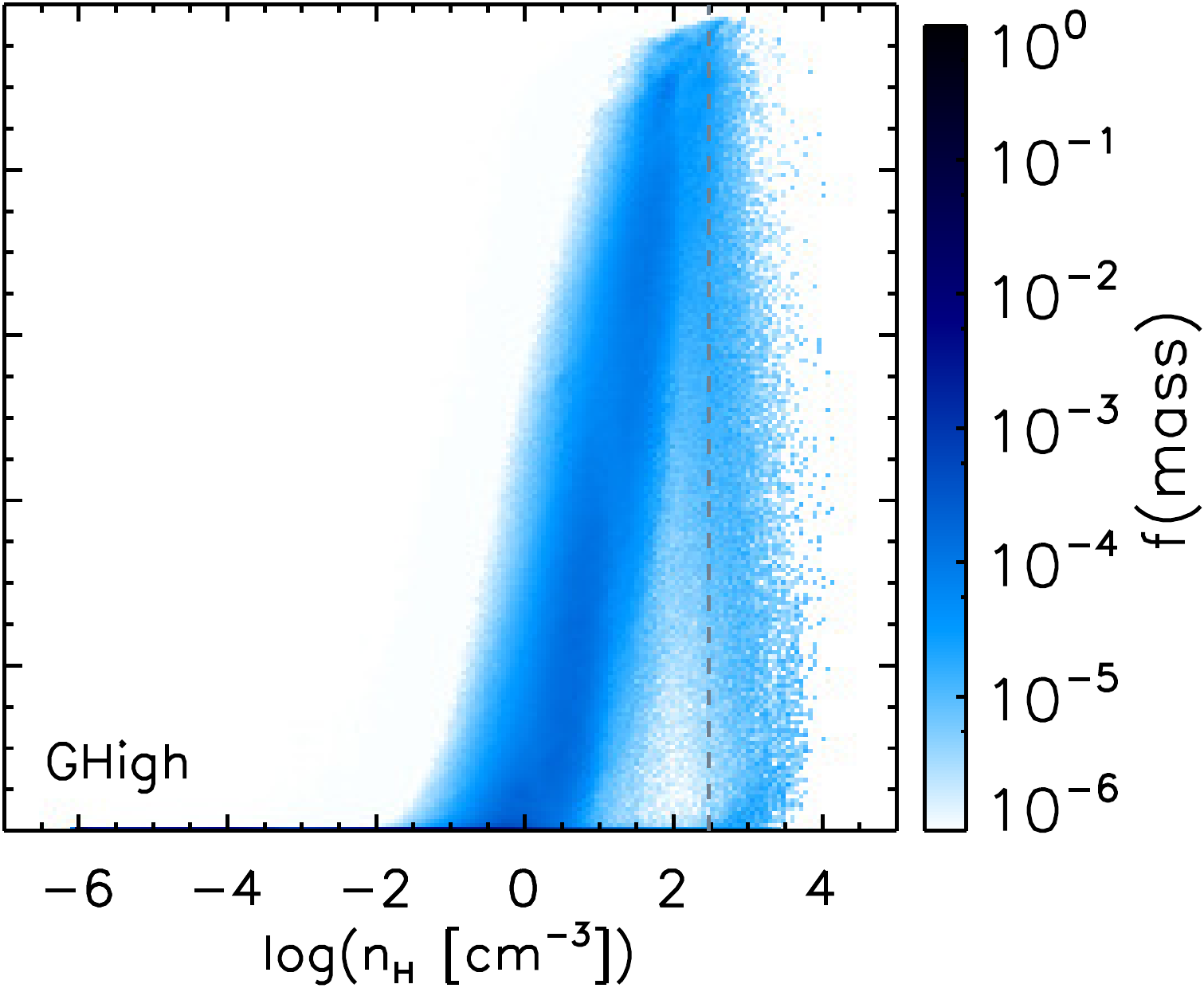}
\end{tabular}
\caption{Phase diagrams of temperature (K) (top row), ionized hydrogen fraction (second from top row), atomic hydrogen fraction (second from bottom row), and twice molecular hydrogen fraction (bottom row) versus hydrogen gas density (\cci), left to right of: GLow, GMed, and GHigh. The star forming region lies right of the dotted line, and in the temperature diagram below the dotted line.} 
\label{fig:phadia}
\end{figure*}

In Fig. \ref{fig:phadia}, we present the phase diagrams for temperature, \hii\ fraction, \hi\ fraction, and \htwo\ fraction versus total hydrogen gas density, and mark our star formation density. Star formation always occurs in our simulation below $3\times10^4$ K, and the density threshold changes depending on the resolution (Table \ref{tbl:galparams}). Every diagram is weighted by gas mass.

Our $n_H$-$T$ diagram shows the characteristic multi-phase nature of these galaxies. The strongest feature shows increasing temperatures at lower densities reaching as high as $10^7$ K at $10^{-6}$ \cci, a constant temperature of about $10^4$ K at intermediate densities around $10^{-4}$ to $10^{-2}$ \cci, and dropping at higher densities to approximately $10^2$ K in GLow, $10^{1.5}$ K in GMed, and a little over 10 K in GHigh. At $\approx10^{-2}$ \cci, the sparse region of gas hotter than $10^4$ K is caused by SN feedback and increases in prevalence with increasing resolution. 

The \hii\ fraction is 1 at the lowest densities and at all resolutions its transition to 0 begins at $10^{-2}$ \cci and ends at $\approx 10^{0}$ \cci. The spread in the intermediate fractions at higher densities increases with increasing resolution, and is probably due to the SN feedback. 

The \hi\ fraction is 0 at the lowest densities but begins to increase with increasing densities following \hii's decrease after $10^{-2}$ \cci due to damping the UV background. However, the \hi\ fraction climaxes at about 0.95, and its rise halts at a little after $10^{0}$ \cci, where it begins to decline. As we pass the star formation threshold there is again an increased in \hi\ at higher fractions. 

The \htwo\ fraction completes the picture. At every resolution, \htwo\ begins forming at densities higher than $10^{-2}$ \cci and its fraction climbs with increasing density until reaching the star formation threshold where it drops back down to 0. Of all the hydrogen species, the \htwo\ abundance is most affected by resolution. In GLow it only goes as high as 0.70, in GMed 0.92, and in GHigh 0.985. 

Comparing the \hi-\htwo\ transition to simulations in other works is difficult because of the number of factors involved. It can vary within the s fame code and model depending on the dust or UV flux \citep{Gnedin2011,Hu2016} or metallicity \citep{Christensen2012,Pallottini2017}. Instead, we can look at the total \htwo\ fraction different codes achieve. As in this work, \citet{Tomassetti2014}'s cosmological simulations do not form 100 per cent molecular gas in their non-equilibrium model. Other simulations by \citet{Gnedin2011,Christensen2012,Hu2016,Capelo2018,Lupi2018} are able to form regions of 100 per cent \htwo, but many of these codes use a clumping factor to enhance \htwo\ formation and account for unresolved substructure. Our philosophy is to not use this clumping factor. Instead, we seek to explain the cause of our simulations' inability to form 100 per cent molecular regions. 

This stems from unresolved Str{\"o}mgren spheres \citep{Stromgren1939}, which is the hot, ionized gas from young stars. This problem was first described in \citet{Rosdahl2015b} in the context of unresolved \hii\ regions. These regions are present in the top row of Fig. \ref{fig:phadia} as diagonal streaks in the star formation region at the highest densities, and is most obvious in GHigh. In the $n_H$-$T$ diagram this is the shorter, higher temperature branch above the main star formation branch, and in the $n_H$-$x_{\mathrm{\hii}}$ diagram this is the increase in \hii\ fraction with decreasing densities that stops at the star formation threshold. In reality the Str{\"o}mgren sphere around a new star should be entirely ionized with a negligible atomic shell embedded in an entirely molecular region, as demonstrated with our code in \citet{Nickerson2018}. However, even in GHigh the resolution goes only down to 6.1 pc, much larger than what many of the Str{\"o}mgren spheres would be around our star particles. This instead produces a cell that is mostly atomic with a smidge of ionized and molecular hydrogen. Unfortunately, given our current star formation model we cannot resolve these small regions, as detailed in \citet{Rosdahl2015b}.

\begin{figure}
\begin{tabular}{l}
\includegraphics[height=0.15\paperheight]{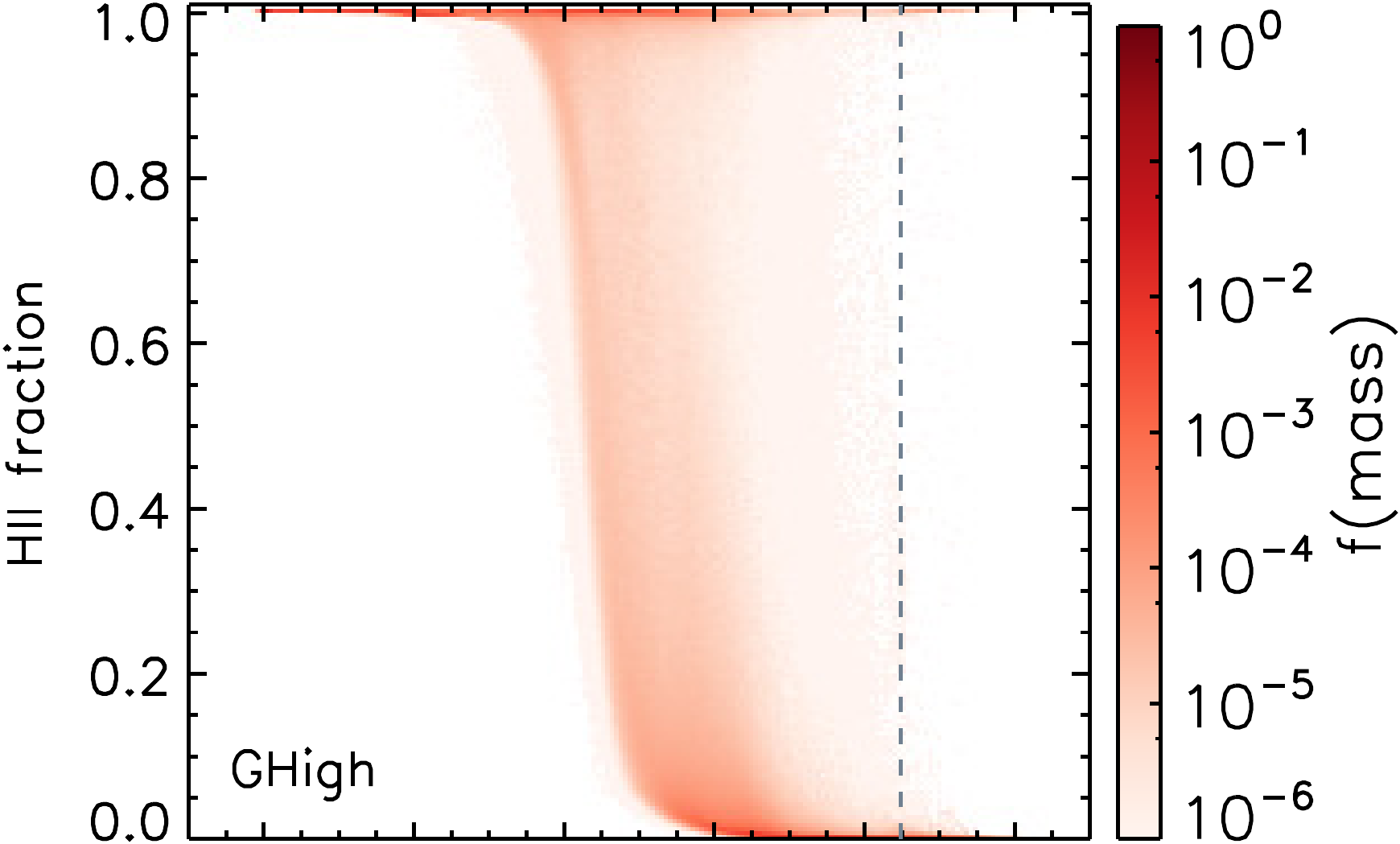} \\
\includegraphics[height=0.15\paperheight]{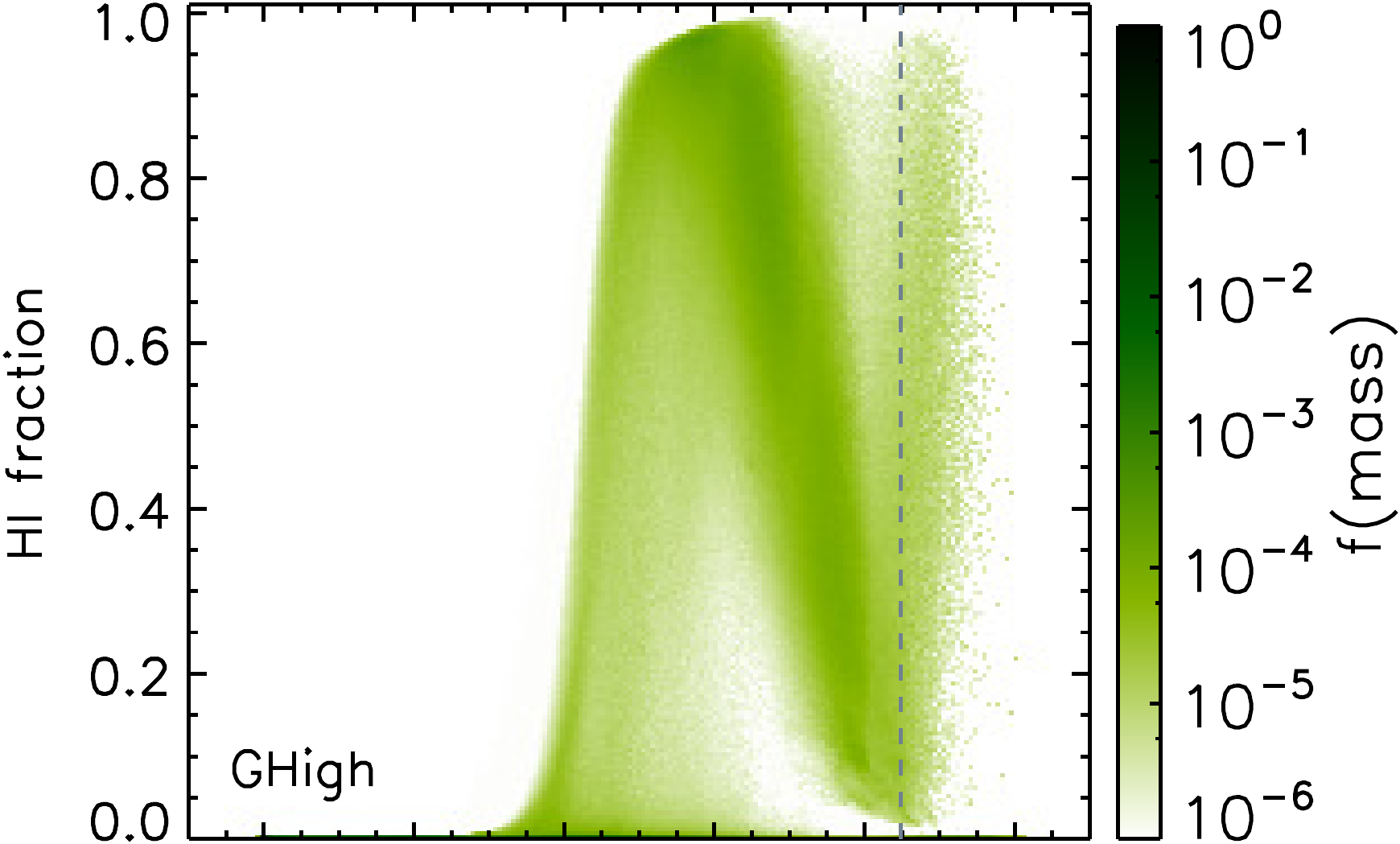} \\
\includegraphics[height=0.177\paperheight]{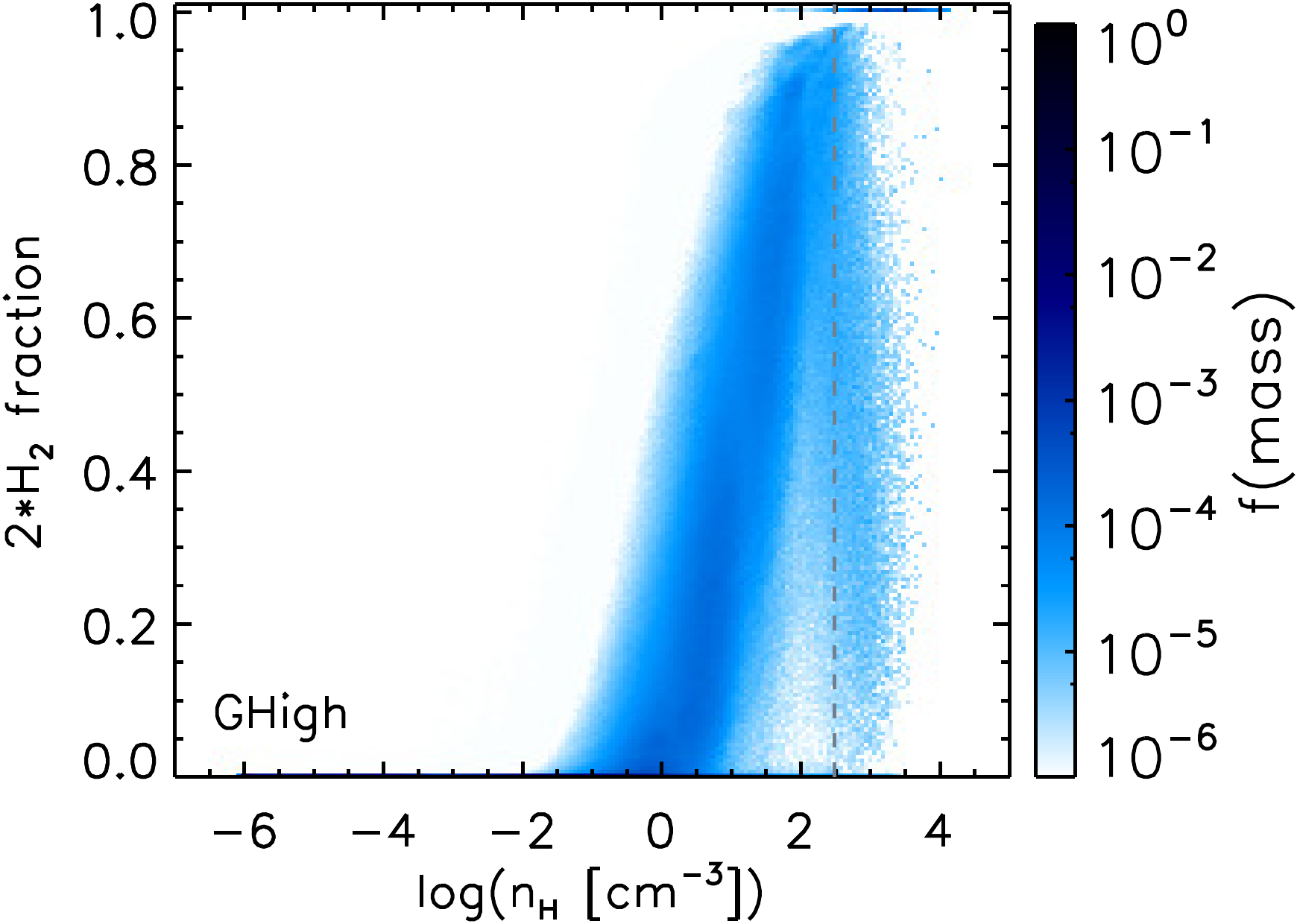}
\end{tabular}
\caption{Post-processed phase diagrams to account for the unresolved Str{\"o}mgren spheres for GHigh, top to bottom: ionized hydrogen fraction, atomic hydrogen fraction, and twice  molecular hydrogen fraction versus hydrogen gas density (\cci). The star forming region lies right of the dotted line.} 
\label{fig:strom}
\end{figure}

We can correct this with post-processing. We consider only cells with stars younger than 10 Myr, the age at which they produce a supernova. Using the SED tables \citep{Bruzual2003} we can calculate the Str{\"o}mgren radius of all stars in a cell, using Equation 62 in \citet{Nickerson2018}. If this radius is less than the distance from the cell centre to the outer corner then in post-processing we split the cell into two regions. The volume inside the radius is considered to be completely \hii, since \hi\ is negligible, while the region outside is considered to be completely \htwo. Fig. \ref{fig:strom} gives the results of this post-processing for GHigh. The high \hii\ fraction above the star formation threshold completely disappears, while there is a lower concentration of \hi, and we obtain a completely molecular fraction at high densities. 

We take this post-processing technique further in applying it to the radial profiles in Fig. \ref{fig:rprof} and the KS relation in Fig. \ref{fig:ksr}. Despite the cells with unresolved Str{\"o}mgren radii being the densest, they actually have a negligible impact on the overall galactic morphologies.  It appears that even though we do not resolve the regions around new stars and the completely molecular region, it has little impact on the overall results.

\subsection{Molecular distribution}
\label{ssec:dist}

As our Introduction highlights, \htwo\ is difficult to observe directly and instead a tracer of even denser gas, CO, is used to convert to \htwo\ abundance. There is observational evidence for dark molecular gas (coined by \citealt{Wolfire2010}) that is not traced by CO \citep{Grenier2005,Burgh2007,Roman-Duval2010}. From the theory side, \citet{Wolfire2010} estimate that about 30 per cent of \htwo\ is dark and \citet{Smith2014} 42 per cent. While our own simulations do not include CO, we can nonetheless quantify the locations of our \htwo\ content.

\begin{figure*}
\begin{tabular}{l|}
\includegraphics[height=0.25\paperheight]{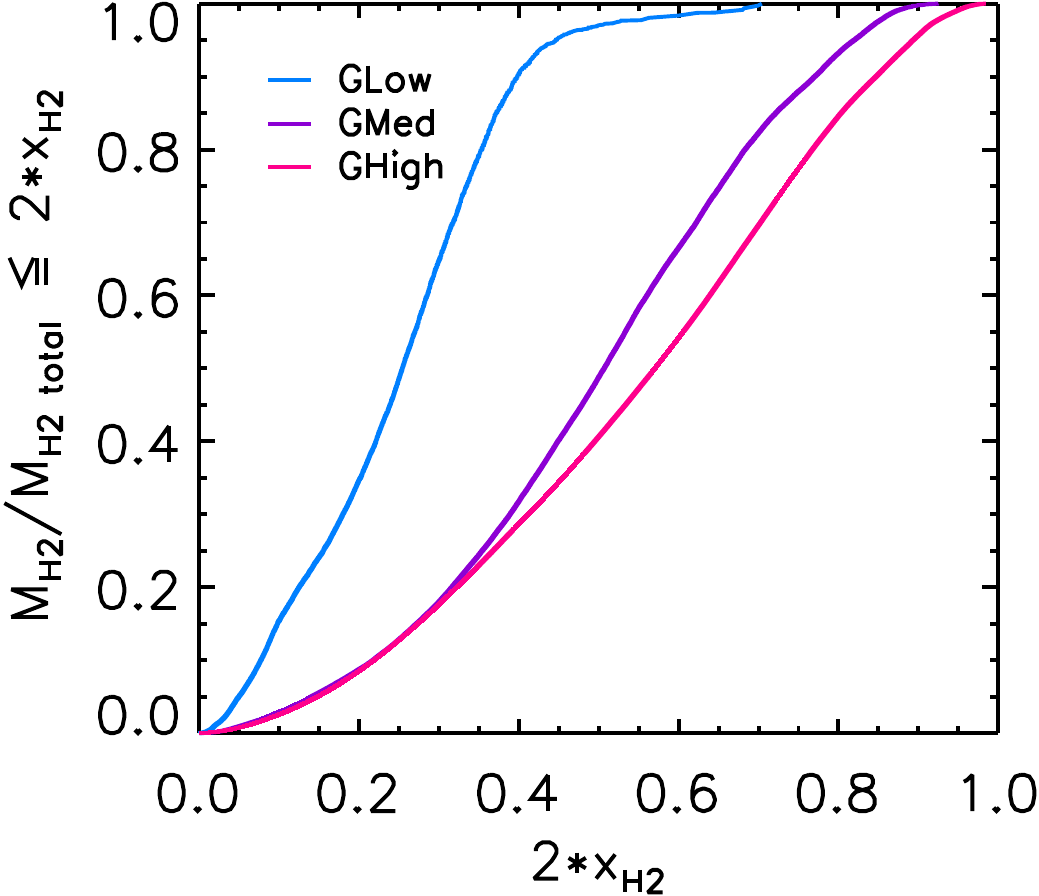} 
\includegraphics[height=0.25\paperheight]{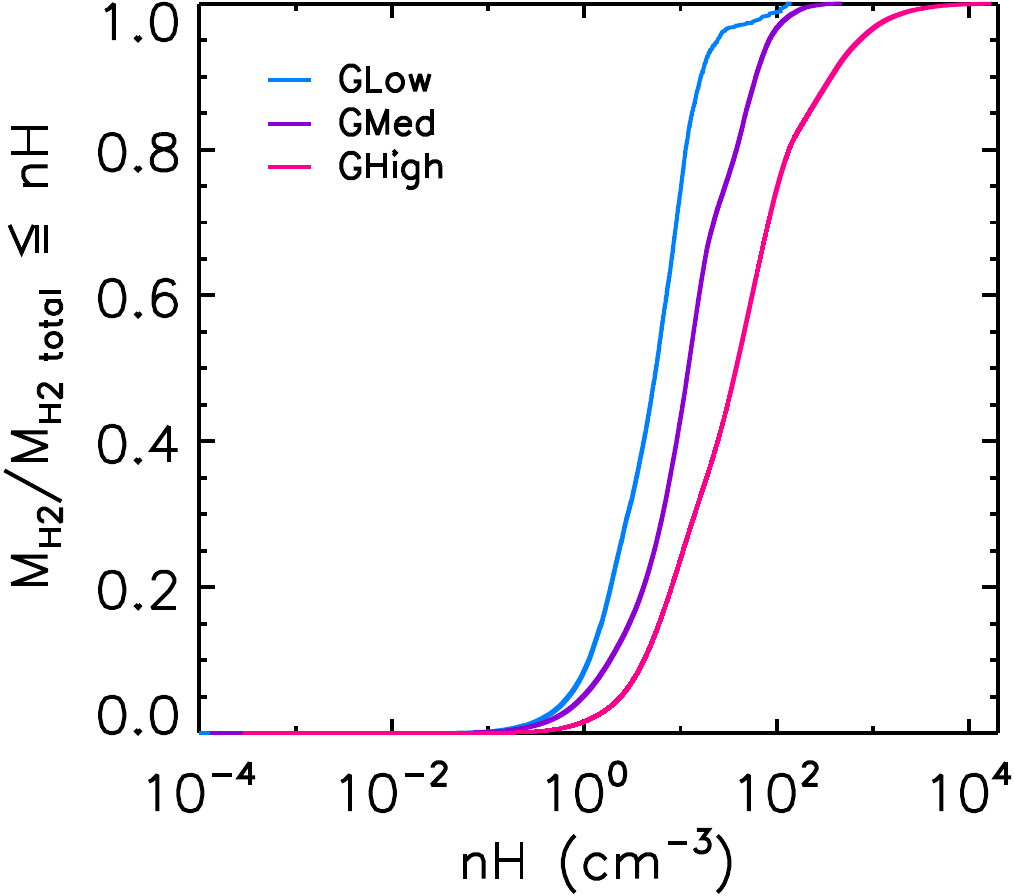} 
\end{tabular}
\caption{The cumulative distribution of \htwo\ mass at each resolution, as a fraction of total \htwo\ mass, left as a function of twice the molecular fraction and right as a function of the total gas density (\cci).}
\label{fig:h2dist}
\end{figure*}

 In Fig. \ref{fig:h2dist}. we show the cumulative mass function of \htwo\ for every cell at all three resolutions versus twice the molecular fraction and the total gas density. Table \ref{tab:mc} quantifies these results, showing the fractions and densities below which 25, 50, 75 and 100 per cent of \htwo\ mass is contained. In the cumulative function for molecular fraction, GLow has drastically less \htwo\ gas, while the GMed curve follows that of GHigh with a little less \htwo. All three resolutions follow the same curve for the cumulative density function, with increasing resolutions reaching higher densities. Table \ref{tab:mc} shows that a significant fraction of \htwo\ exists in mixed regions, with half in gas that is only 25.4, 50.7, and 57.0 per cent \htwo\ or less for GLow, GMed and GHigh respectively. Half of the \htwo\ gas in regions that are below 5.57, 12.0, and 37.8 \cci\ for GLow, GMed, and GHigh. Clearly, a significant fraction of our \htwo\ gas is in regions less dense than the average molecular cloud (see Figure \ref{fig:clhist}) and is significantly mixed with \hi. The phase diagrams (Fig. \ref{fig:phadia}) emphasize this as well, albeit in a more qualitative than quantitative fashion.

\begin{table}
\begin{tabular}{cc|c|ccc}
\hline
\hline
&$M_{\text{\htwo total}}$&GLow&GMed&GHigh\\
\hline
$2x_{\text{\htwo}}$ & $\le 25\%$&0.155&0.354&0.368\\
& $\le 50\%$&0.254&0.507&0.570\\
&$\le 75\%$&0.334&0.652&0.734\\
&$\le 100\%$&0.704&0.925&0.985\\
\hline
$n_H$ (\cci) & $\le 25\%$&2.30&5.34&10.6\\
& $\le 50\%$&5.57&12.0&37.8\\
& $\le 75\%$&10.2&29.1&102\\
&$\le 100\%$&143&476&$1.70\times10^4$\\
\hline
\hline
\end{tabular}
\caption{Distribution of total \htwo\ gas corresponding to Fig. \ref{fig:h2dist}. Columns left to right: quantity type, either twice the \htwo\ fraction or total gas density (\cci); percentage of total \htwo\ mass at which the quantity is less than or equal to the values in the three columns to the right; GLow; GMed; and GHigh.}
\label{tab:mc}
\end{table}

\section{Molecular cloud analysis}
\label{sec:mc}

Having looked at the overall properties of our galaxy, we now examine the molecular clouds within it. For this analysis we use GHigh because only it has the resolution to study the clouds in satisfactory detail, the typical cloud density being 100 \cci \citep{Dobbs2013b}. GLow and GMed, as seen in Fig. \ref{fig:phadia} are unable to form high enough gas densities in high enough quantities to be comparable to molecular clouds.

\subsection{Clump finding and cloud properties}
\label{ssec:cf}

We identify the clouds using \ramses's native clump finder Parallel Hierarchical Watershed (PHEW) \citep{Bleuler2014}, but base our search on \htwo\ density as opposed to total gas density. However, using total gas density does not change the results. We set the density threshold as 100 \cci\ for a cloud's peak cell \htwo\ density, which is the typical molecular cloud density \citep{Dobbs2013b}, with a relevance threshold of 10. This density threshold for simulated molecular clouds is introduced in \citet{Tasker2009}. The saddle threshold above which two clumps are merged is 1000 \cci. We only consider clumps with at least 10 grid cells. The \htwo\ mass threshold is 10 \msun, a little higher than the lower limit observed in the Milky Way \citep{Miville-Deschenes2017}, though as we will see in Fig. \ref{fig:clcume} our least massive cloud is actually on the order of $10^4$ \msun\ since the cell number is the more stringent factor. Our clump finder PHEW works in position-position-position (PPP) space, which is native to simulations, while observations find clouds in position-position-velocity (PPV) space because the light of sight coordinate can only be obtained through velocity. \citet{Pan2015} show that both PPP and PPV analysis on the same simulation do produce similar cloud properties and structures.

We consider only the molecular component when summing up each cloud's mass. The clump finder returns the volume of each cloud as summed over the individual cells that comprise it, and we calculate the radius by approximating each cloud as a sphere. We find the surface density via: 
\begin{equation}
\Sigma_C = M_C / (\pi R_C^2)
\label{fig:sd}
\end{equation}
where $\Sigma_C$ is the surface density,  $M_{C}$ is the total \htwo\ mass of the cloud, and $R_C$ is the cloud radius.

A cloud's velocity dispersion is an important quantity, comprising the turbulent motion and the thermal components. For comparison we consider our clouds as observed face-on and use only the z-direction perpendicular to the galactic plane for calculating the turbulent velocity dispersion, $\sigma_{v,\rm{turb}}$:
\begin{equation}
\sigma_{v,\rm{turb}}=\sqrt{\frac{\sum_i m_{\htsub,i} (v_{z,i}-\bar{v}_z)^2}{M_C}},
\label{eqn:turb}
\end{equation}
where $\bar{v}_z$ is the \htwo-mass weighted mean velocity in the $z$ direction, $i$ is the index of each cell in a cloud, $v_{z,i}$ is the $z$ velocity of that cell, and $m_{\htsub,i}$ is the \htwo\ mass in that cell. For the thermal component, we consider the sound speed, $\sigma_{v,\rm{therm}}$:
\begin{equation}
\sigma_{v,\rm{therm}}=\sqrt{\frac{\gamma k_B T_C} {\mu m_H}},
\label{eqn:sound}
\end{equation}
where $\gamma=5/3$ is the heat capacity ratio for this simulation, $k_B$ is the Boltzmann constant, $T_C$ is the \htwo\ mass-weighted average cloud temperature, $\mu$ is the mean molecular mass, and $m_H$ is the mass of a hydrogen atom. Combining the two components, we acquire the total velocity dispersion, $\sigma_v$:
\begin{equation}
\sigma_{v}=\sqrt{\sigma_{v,\rm{turb}}^2 + \sigma_{v,\rm{therm}}^2 }.
\label{eqn:sigv}
\end{equation}
We note that, as expected for cold molecular clouds, the turbulent component dominates. 

The dimensionless virial parameter $\alpha_{vir}$ \citep{Bertoldi1992} encapsulates the balance between gravitational and kinetic energy in a cloud: 
\begin{equation}
\alpha_{vir}=\frac{2K_C}{|W_C|}=\frac{5\sigma_v^2R_C}{GM_C}
\end{equation}
where $K_C$ is the kinetic energy of the cloud, $W_C$ is the potential energy of the cloud, and $G$ is the gravitational constant. Clouds with $\alpha_{vir} \approx 1$ are considered to be in virial equilibrium, while clouds with $\alpha_{vir} >> 1$ need to be supported by internal pressure against the surrounding gas or are otherwise unbound and transient. $\alpha_{vir} << 1$ means that a cloud requires a magnetic field support in order to maintain virial equilibrium \citep{Bertoldi1992} though we do not include magnetic fields in these simulations.

\subsection{Cloud maps}
\label{ssec:cmap}

\begin{figure*}
\includegraphics[width=0.95\textwidth]{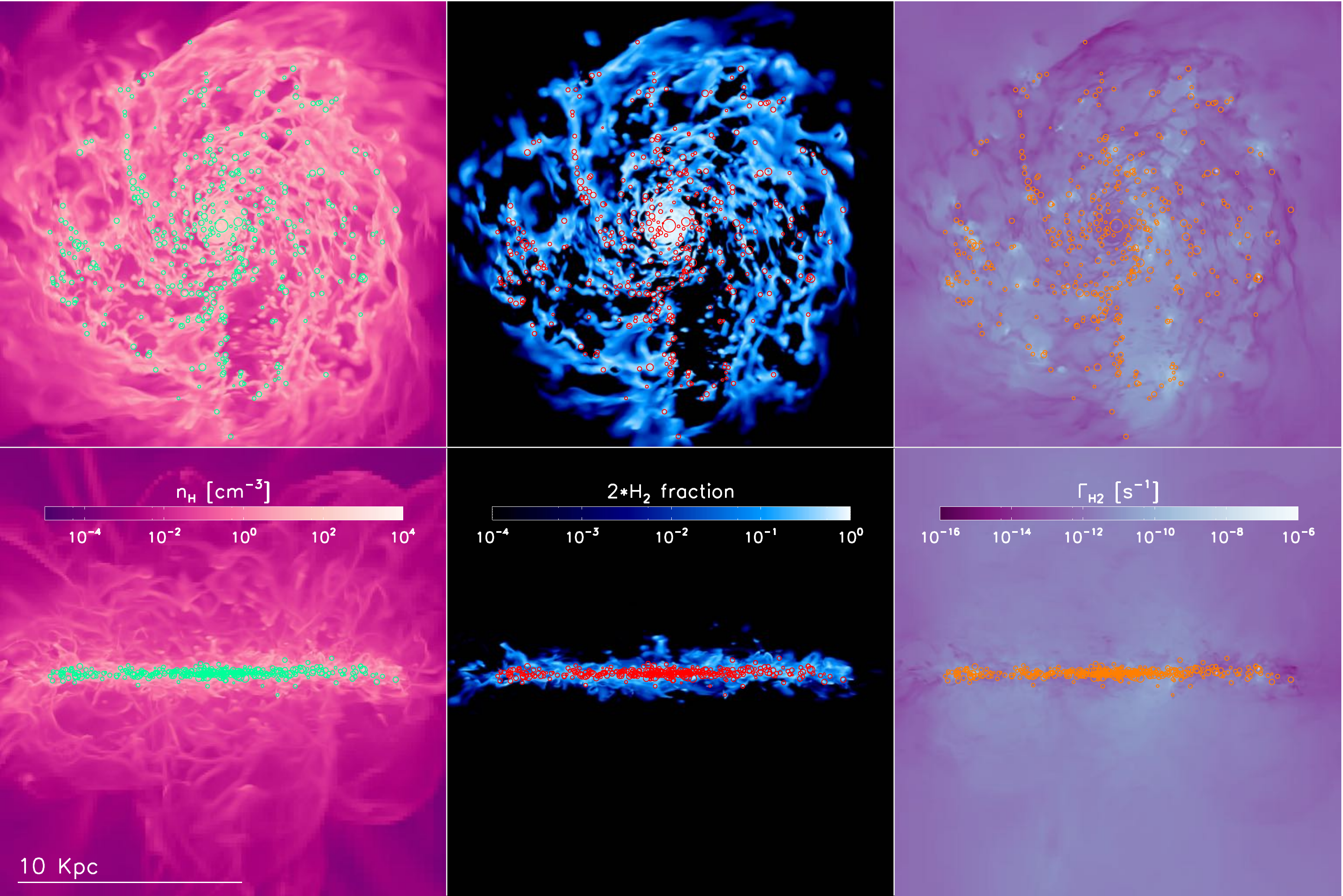}
\caption{Positions of our clouds within GHigh, represented by circles drawn at four times their radii, top row face-on and bottom row side-on, left to right, each property mass weighted: total gas density (\cci), \htwo\ fraction, and total photodissociation and ionization rate of \htwo\ (s$^{-1}$). The molecular clouds are coloured left to right: green, red, and orange to contrast each map.}
\label{fig:clmap}
\end{figure*}

Fig. \ref{fig:clmap} shows zoomed-in maps of our galaxy, both face-on and side-on, for total gas density, molecular hydrogen fraction, and total photodissociation and ionization rate for \htwo\ superimposed with the locations of our molecular clouds. We are unable to resolve clouds closest to the galactic centre but we resolve them for the remainder of the disc. They trace out the spiral arms, and the inter-arm region contains significantly fewer clouds. 

Our clouds are part of a continuum of wider high \htwo\ regions. Looking at the clouds side-on, they are mostly confined to the disc. The clouds are nearly entirely molecular with the average 96 per cent molecular, and the range is 93 to 97 per cent. The average \hi\ fraction is 4 per cent and the average \hii\ fraction is 0.04 per cent. 

\begin{figure}
\includegraphics[width=0.75\columnwidth]{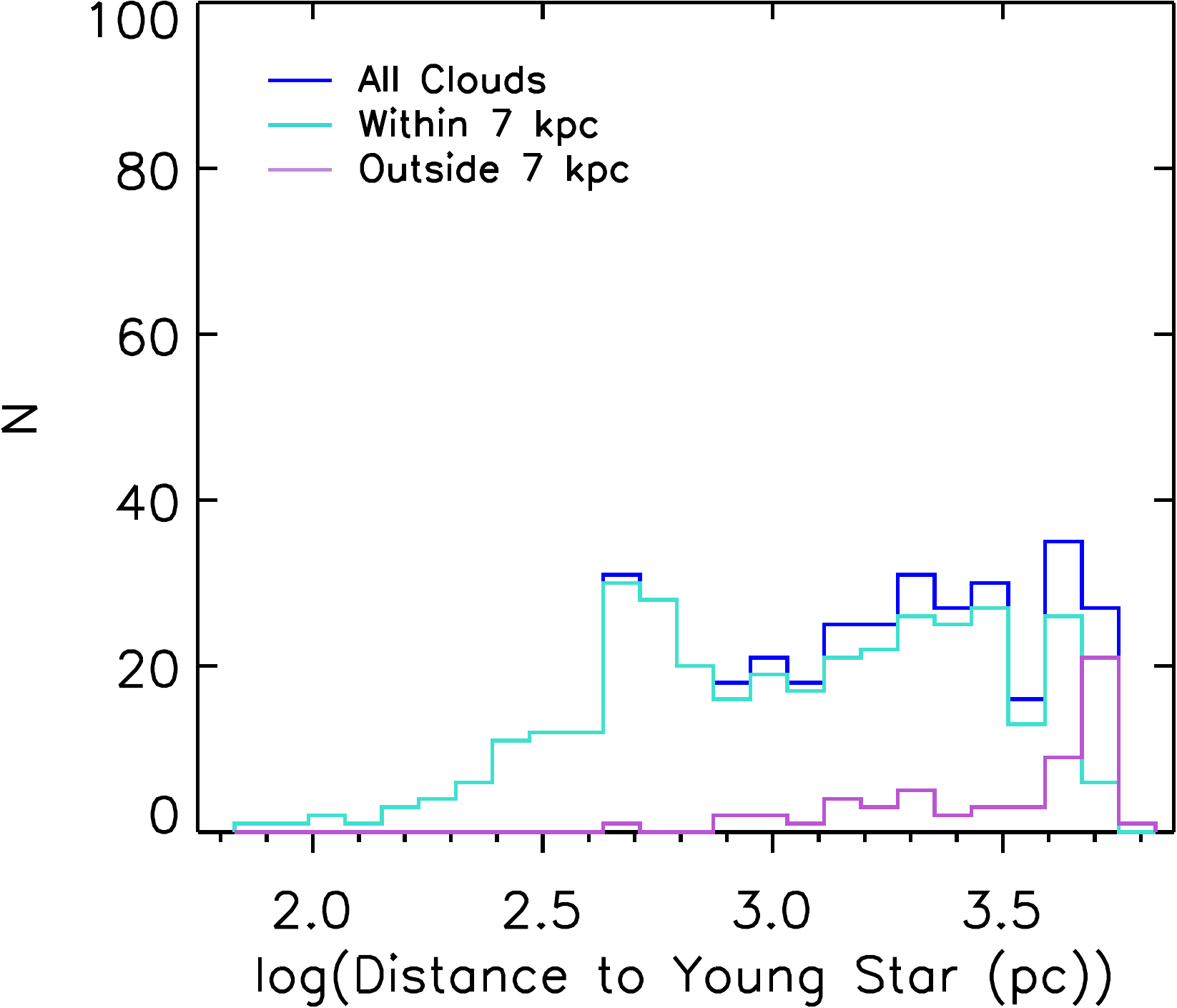}
\caption{Histogram of the distance of the nearest young ($\leq 10$ Myr) star to a cloud boundary (pc). $N$ is the total number of clouds in each bin. The clouds are separated into total cloud population (blue), inner clouds (turquoise), and outer clouds (plum) as explained in Section \ref{ssec:cprops}}
\label{fig:sdist}
\end{figure}

When comparing the clouds to the photon map, which traces the energy from young stars, it is clear that most of the clouds do not match these energy sources. We need to see, however, if any of the clouds that do align with the energy sources truly do, or if this is a projection effect. We calculate the nearest distance from each cloud to a young ($\leq$ 10 Myr) star, and find that no young stars are within cloud radii. Fig. \ref{fig:sdist} shows a histogram of the distance of each young star to its nearest cloud's outer radius. This histogram is further divided into two cloud populations, inner and outer disc, for reasons we will elaborate on in Section \ref{ssec:cprops}. A negative value would imply that the star is inside the cloud, and this is satisfied by none of the clouds, the nearest being 68 pc away. This would suggest that once a cloud forms a star, the radiation quickly dissociates its high density gas.

A further consequence of this is that we cannot produce SFR relations for our molecular clouds. However, \citet{Khoperskov2017} examine the KS relation on several scales for the same simulated galaxy, from 200 to 4 pc, and find that the KS relation breaks down 50 pc, which our lack of young stars in clouds supports. 

\subsection{Cloud observational comparison}
\label{ssec:cprops}

\begin{figure}
\includegraphics[width=0.9\columnwidth]{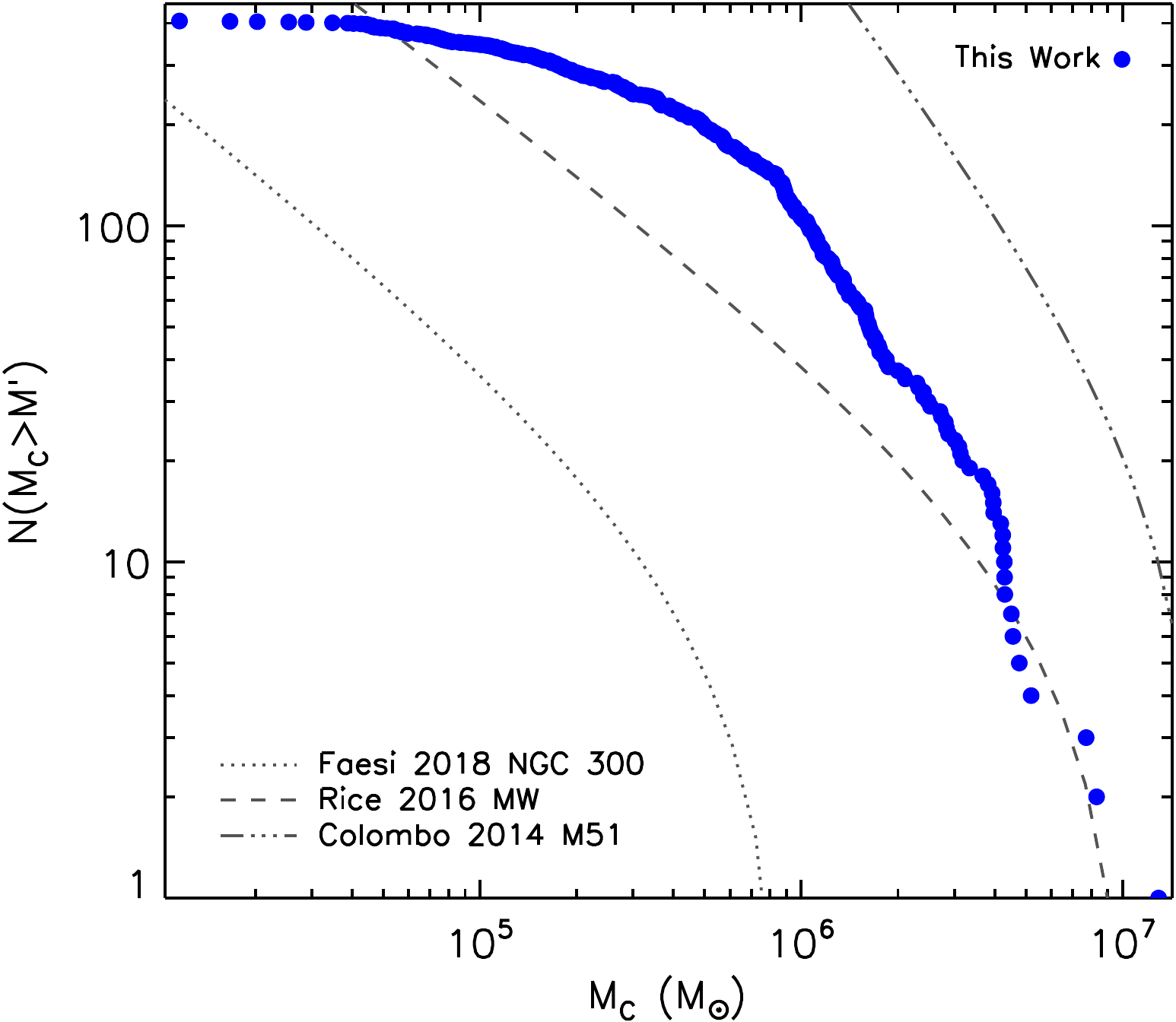}
\caption{Cumulative function of our molecular cloud masses (\msun, blue dots), compared to functions for NGC 300 \citep{Faesi2018} (dotted line), the Milky Way \citep{Rice2016} (dashed line), and M51 \citep{Colombo2014} (dash-dotted line).}
\label{fig:clcume}
\end{figure}

Cumulative mass functions for observed cloud masses are fit either with a power law or truncated power law \citep{Williams1997}. Most molecular cloud properties are remarkably uniform across different galaxies except for the cumulative mass function \citep{Rosolowsky2005}. We present the cumulative mass function of our clouds in Fig. \ref{fig:clcume} and compare our simulated data to the fitted functions from three different galaxies: NGC300 \citep{Faesi2018}, the Milky Way \citep{Rice2016}, and M51 \citep{Colombo2014}. Being the least massive galaxy, NGC300 harbours fewer and less massive clouds. Our galaxy, being a Milky Way analog, does show a similar distribution of high mass ($10^7$\msun) clouds to the Milky Way, but we have a surplus of intermediate-mass clouds ($10^6$\msun). The cumulative mass function of M51 is the most massive shown here. Clouds as massive as $10^8$\msun\ are observed in our Galactic centre \citep{Oka2001}, but these massive clouds are missing from the surveys we show here and our own simulation. At the other end, we have fewer low-mass clouds because of our limit of at least ten cells per cloud.

All of our clouds fall within the central 10 kpc radius of our galactic disc, and comprise 25 per cent of molecular mass in the disc. This is the same fraction given by \citep{Rice2016} for the Milky Way, but lower than the 54 per cent given by \citep{Colombo2014} for M51. This matches our findings in Table \ref{tab:mc}, where 75 per cent of \htwo\ in GHigh is of density 102 \cci or lower, close to our threshold in the clump finder.

Next, we seek to compare our cloud population to the Milky Way cloud catalogue compiled by \citet{Miville-Deschenes2017} from data by \citet{Dame2001}. They find that the distributions of cloud properties differ between the inner and outer disc of the Milky Way, where the clouds in the outer disc have lower densities and lower masses, but a higher virial parameter. \citet{Colombo2014} further subdivided M51 into inner and outer spiral arms and inter-arms regions, but this detailed analysis goes beyond our resolution. M51's clouds have higher densities and masses in the central galaxy and the arms, while these properties are lowest in the inter-arm regions. 

\begin{figure*}
\begin{tabular}{l|}
\includegraphics[height=0.17\paperheight]{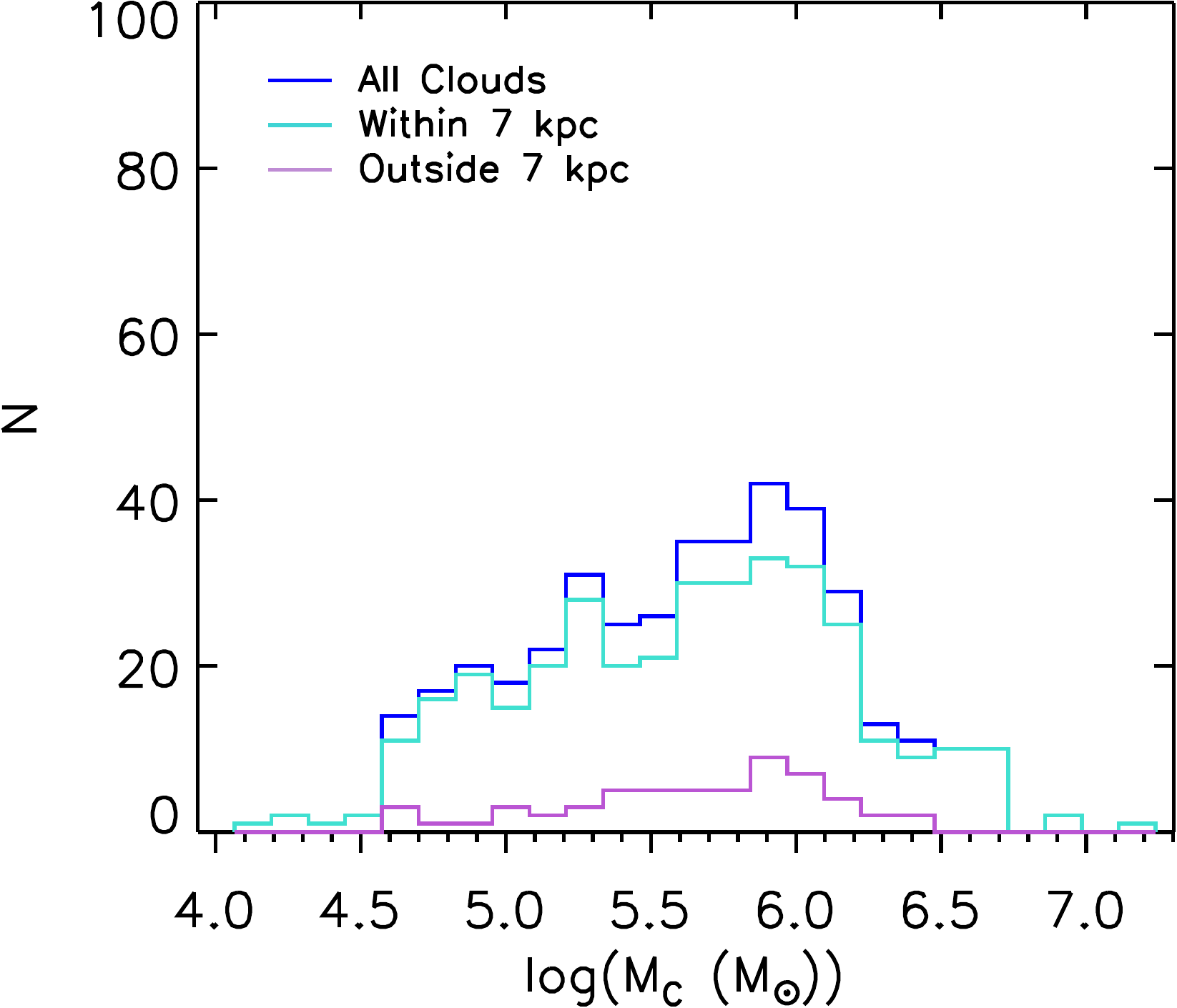} 
\includegraphics[height=0.17\paperheight]{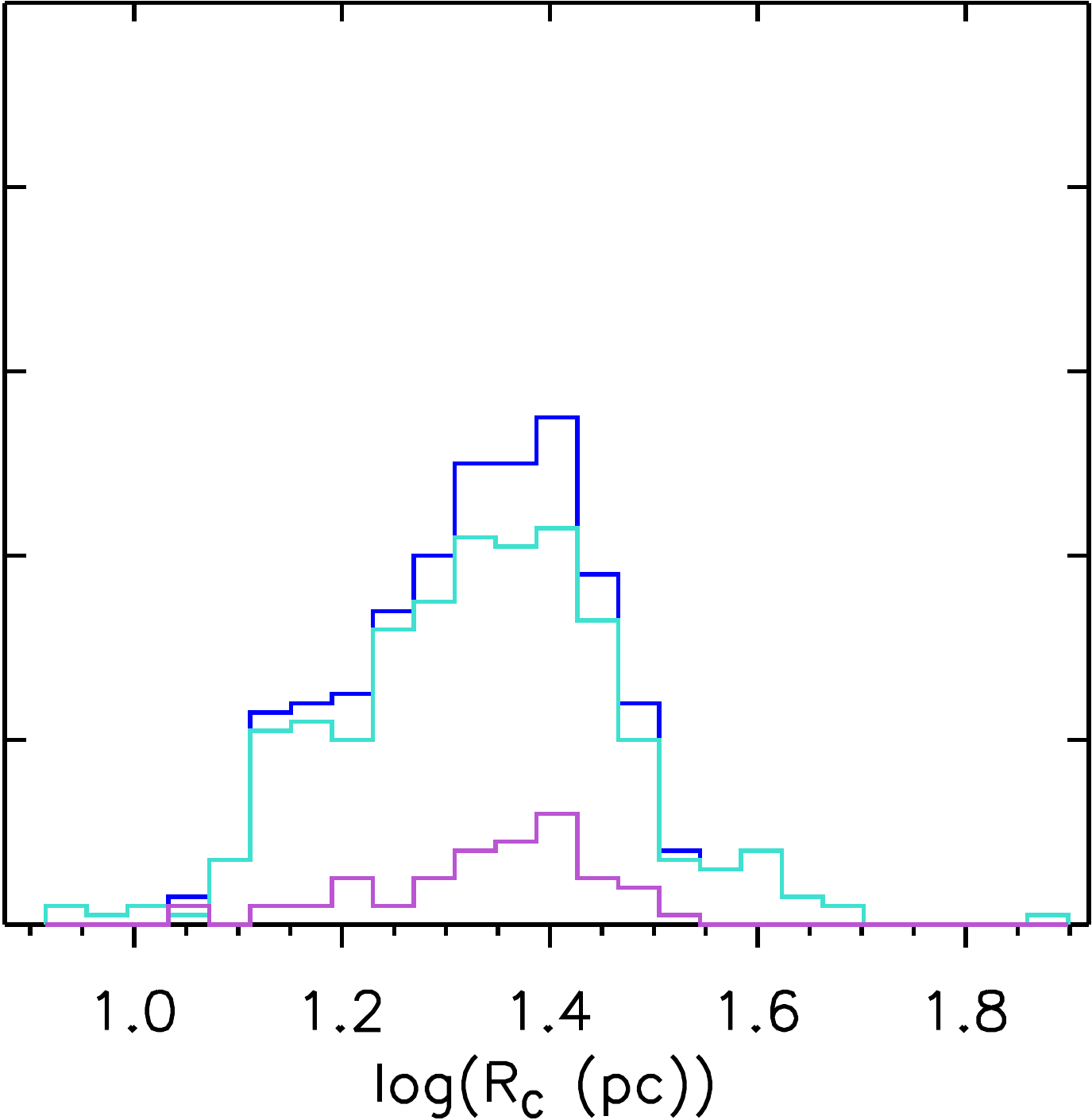} 
\includegraphics[height=0.17\paperheight]{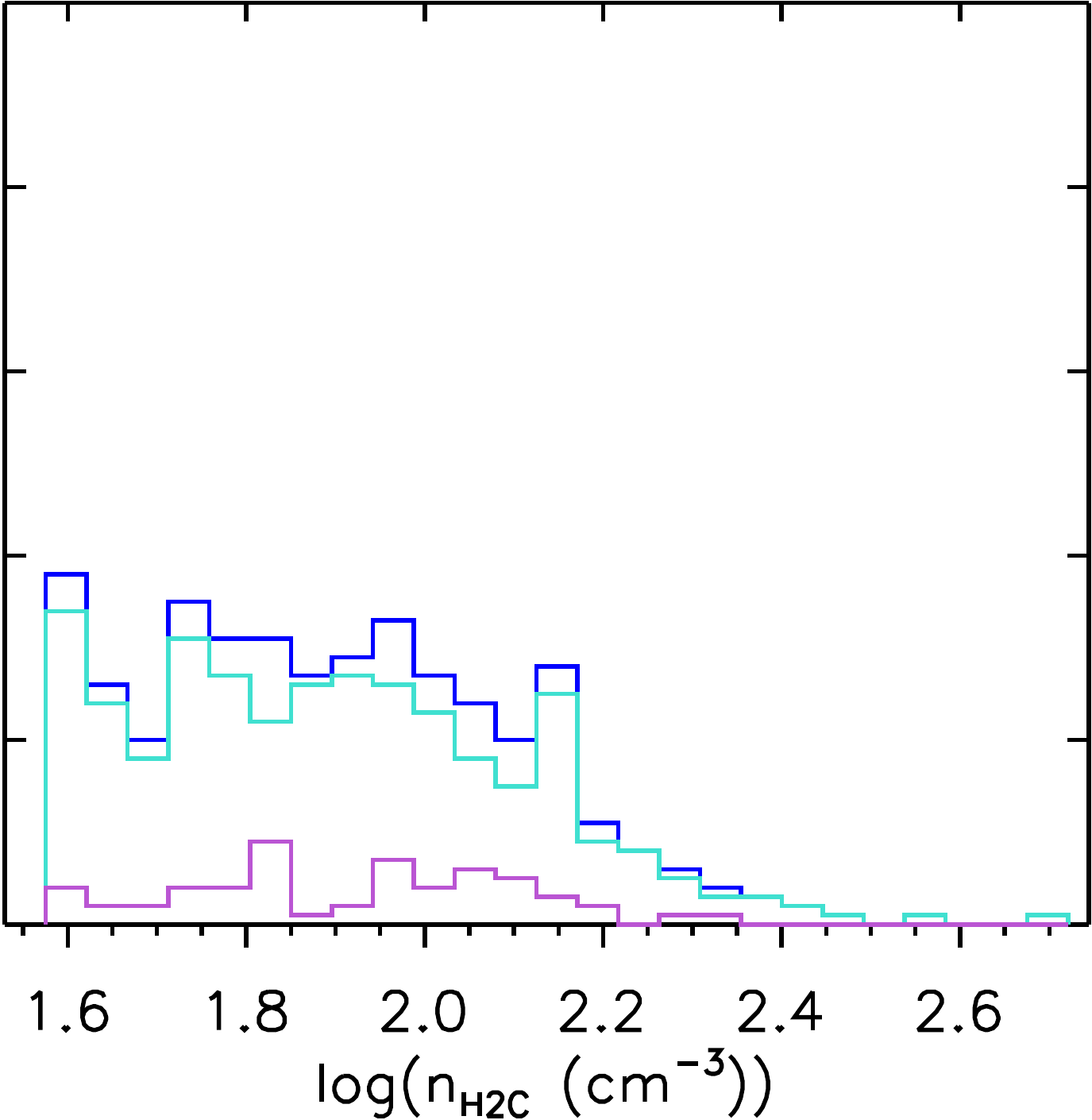} \vspace{0.5 cm}\\
\includegraphics[height=0.17\paperheight]{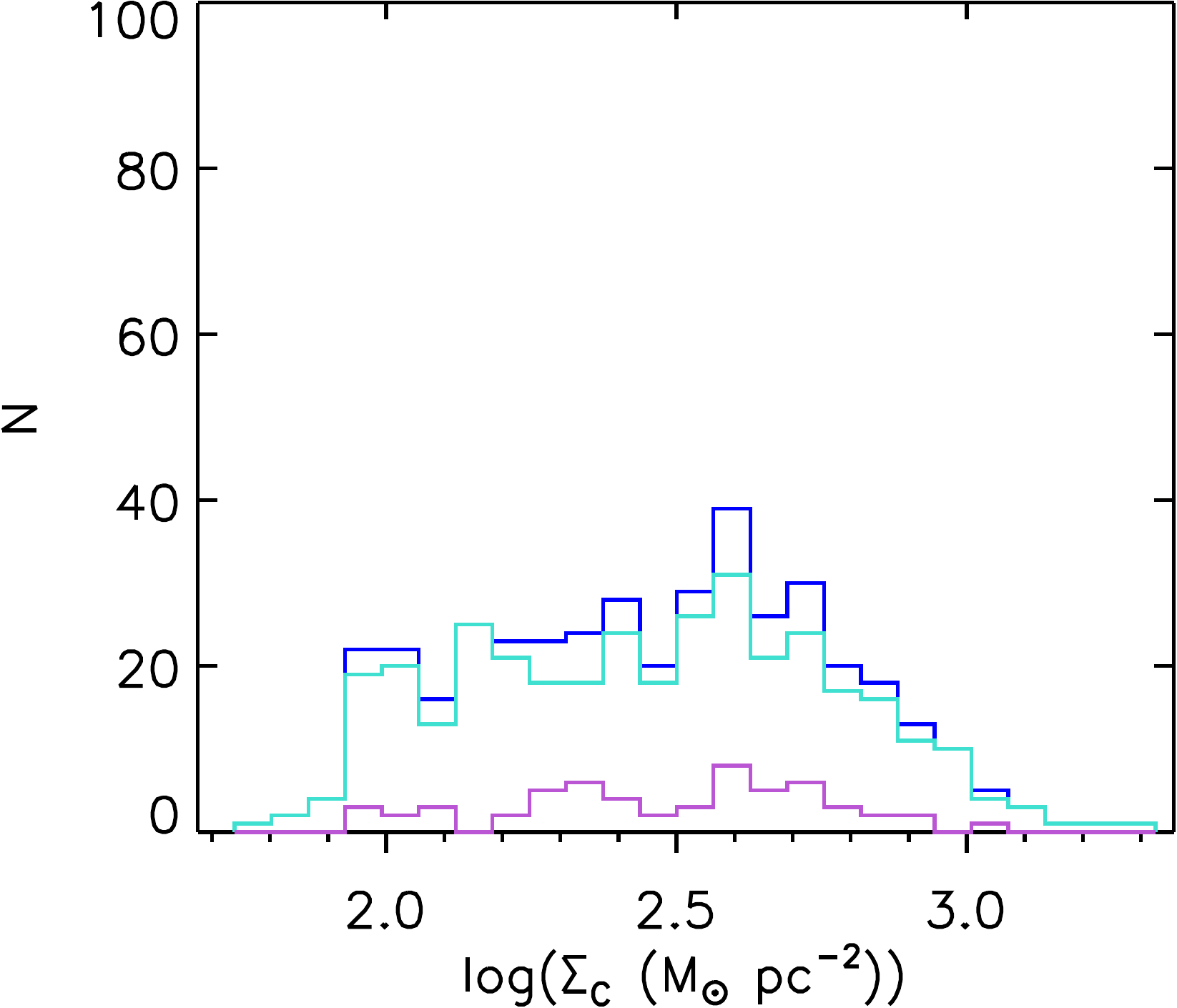} 
\includegraphics[height=0.17\paperheight]{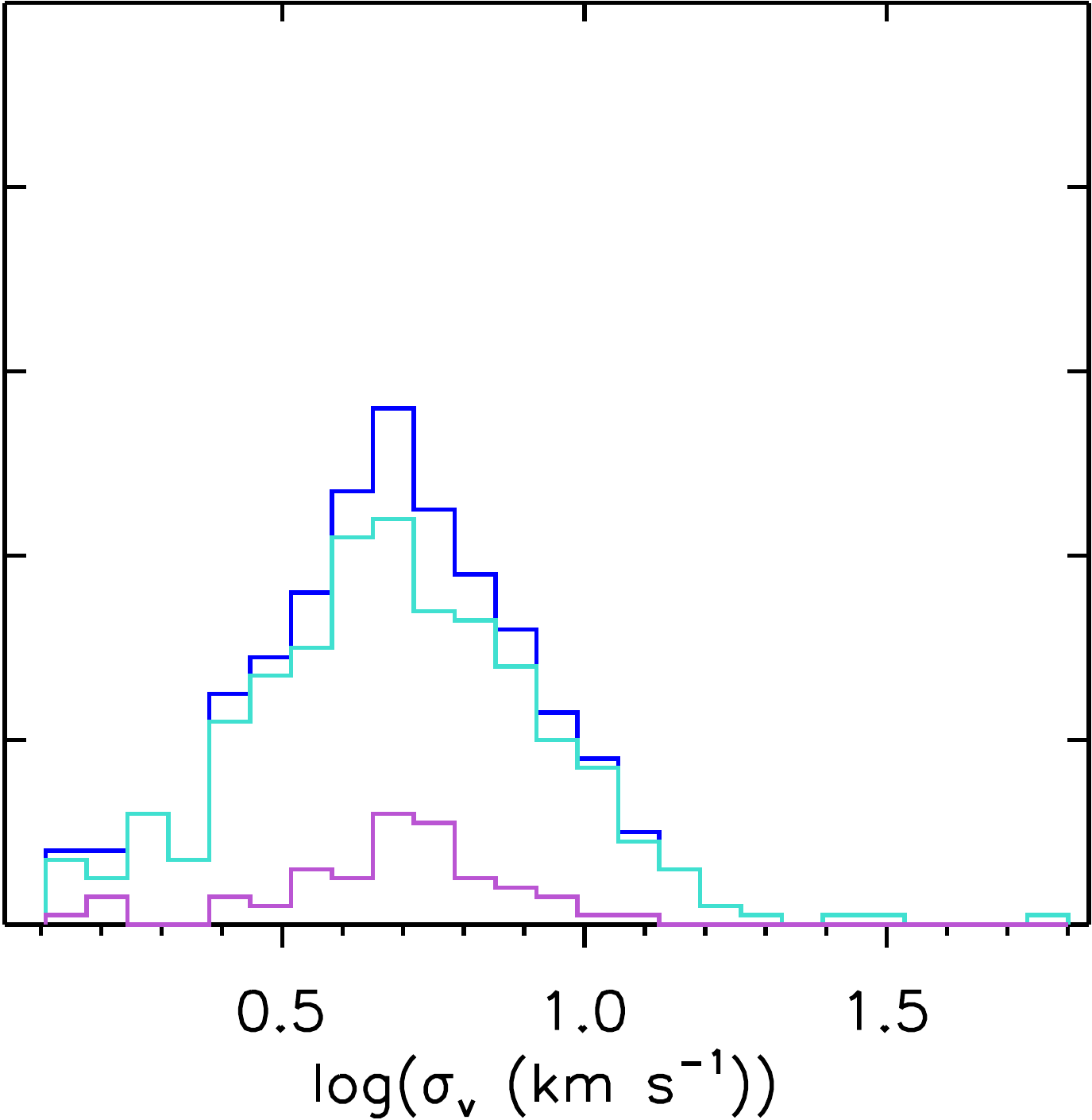} 
\includegraphics[height=0.17\paperheight]{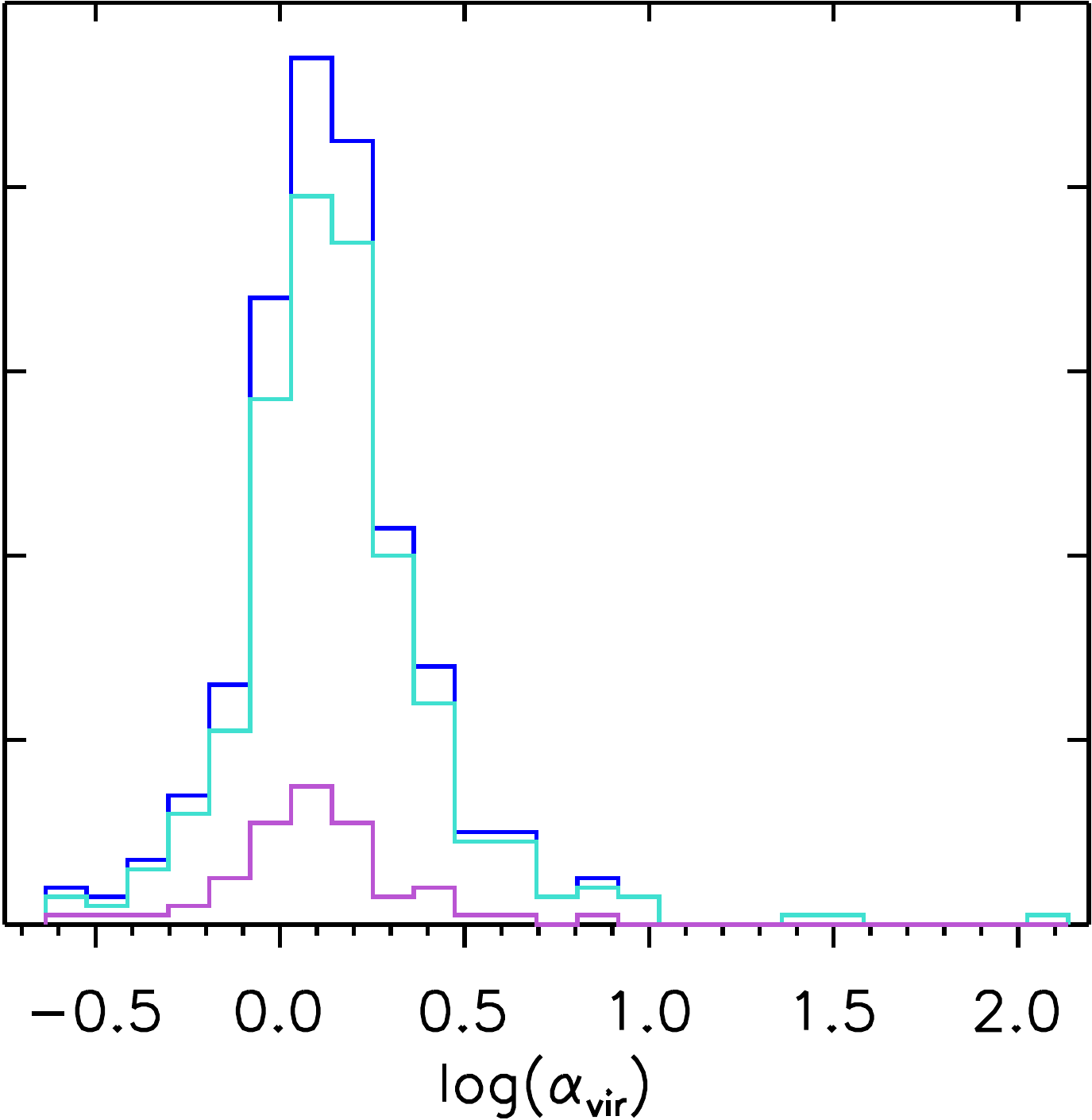}
\end{tabular}
\caption{Histograms for our total cloud population (blue), inner (turquoise), and outer clouds (plum), left to right, top to bottom: mass (\msun), radius (pc), volume density (\cci), surface density (\msun pc$^{-2}$), velocity dispersion (\kms), and virial parameter. $N$ is the total clouds in each bin.} 
\label{fig:clhist}
\end{figure*}

We show histograms in Fig. \ref{fig:clhist}, with the cloud population divided into the inner disc (inside 7 kpc) and the outer disc (outside 7 kpc) where the \htwo\ and SFR fall off in Fig. \ref{fig:rprof}. Each histogram contains 25 bins evenly spaced logarithmically. The cloud properties are: mass, radius, volume density, surface density, velocity dispersion, and the virial parameter. Our clouds range from a little over $10^7$ \msun\ to a few $10^4$ \msun, peaking at $10^6$ \msun, and 10 to 70 pc in radius, peaking at 25 pc. We also show the volume density of \htwo\ in which the distribution favours lower densities of around 40 \cci, and ranges from about 40 to to 400 \cci . The surface density peaks at about 400  \msun pc$^{-2}$ toward the more massive end and ranges from 60 to 2000 \msun pc$^{-2}$. Our velocity dispersion histogram is even more peaked at 5 \kms, ranging from 1.5 to 60 \kms. Finally, we see that the virial parameter peaks at 1.25, which is slightly over virial equilibrium between internal kinetic and gravitational forces. The virial parameter is the most peaked of all our distributions. However, clouds with $\alpha_{vir} \approx 2$ are still considered to be marginally gravitationally bound \citep{Dobbs2013b}.

When we consider the clouds as divided between inner and outer discs, a different story emerges from that of \citet{Miville-Deschenes2017}. We see very little regional distinctions. There are fewer clouds in the outer disc, given that it also covers a smaller surface area and lower \htwo\ density. The mass, radius, volume, and surface density distributions are the same between the two populations. The outer clouds do not have extremes on either end, but this may be due to small-numbers statistics. Both inner and outer cloud distributions peak in the same locations. We do not see the drastic bi-modality as in \citet{Miville-Deschenes2017}, though one explanation is that we are not able to resolve the low density clouds that they can in their observations.

\begin{figure*}
\begin{tabular}{c|c}
\includegraphics[height=0.18\paperheight]{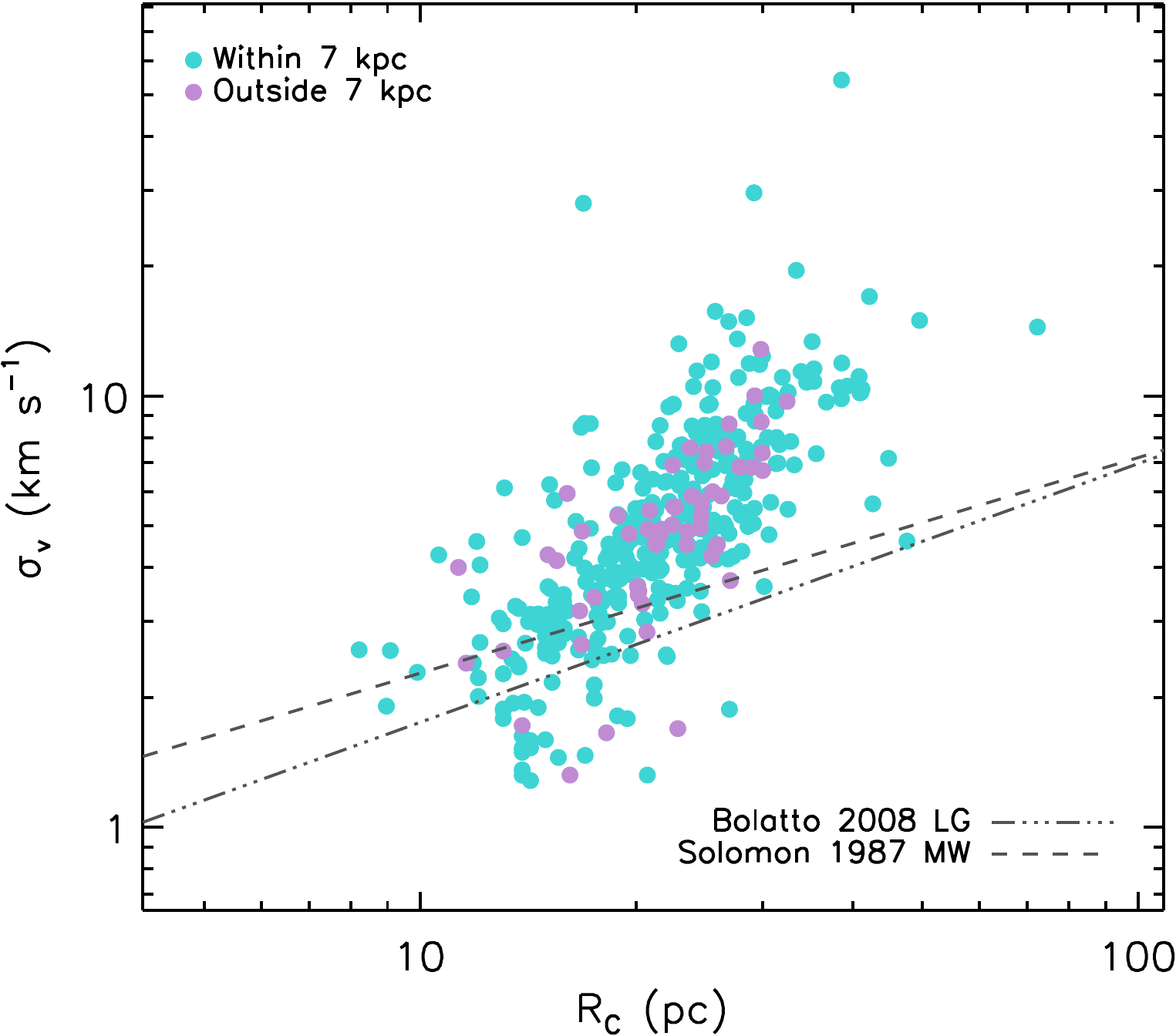} \hspace{0.5 cm}
\includegraphics[height=0.18\paperheight]{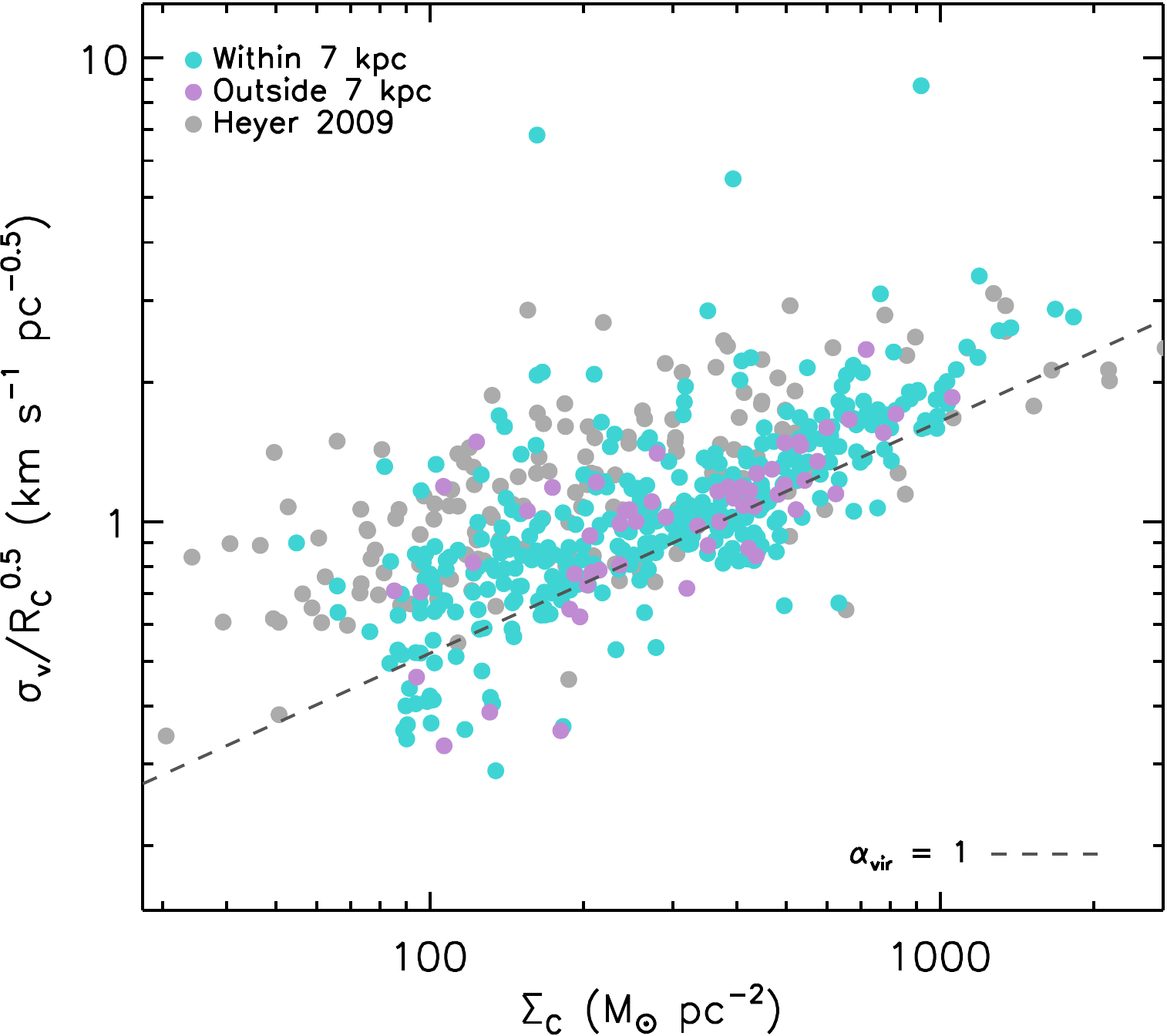} \vspace{0.5 cm} \\
\includegraphics[height=0.18\paperheight]{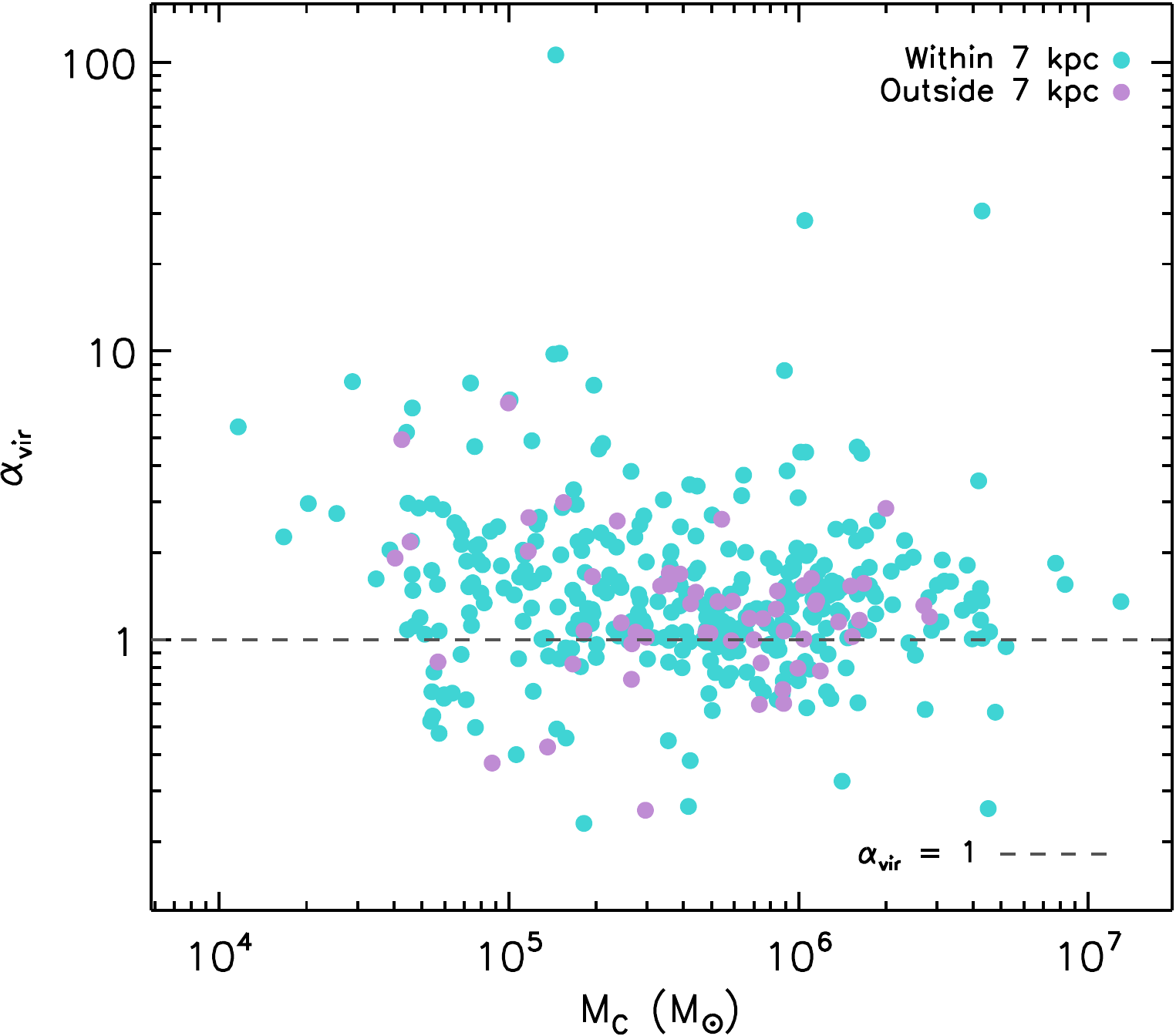} \hspace{0.5 cm}
\includegraphics[height=0.18\paperheight]{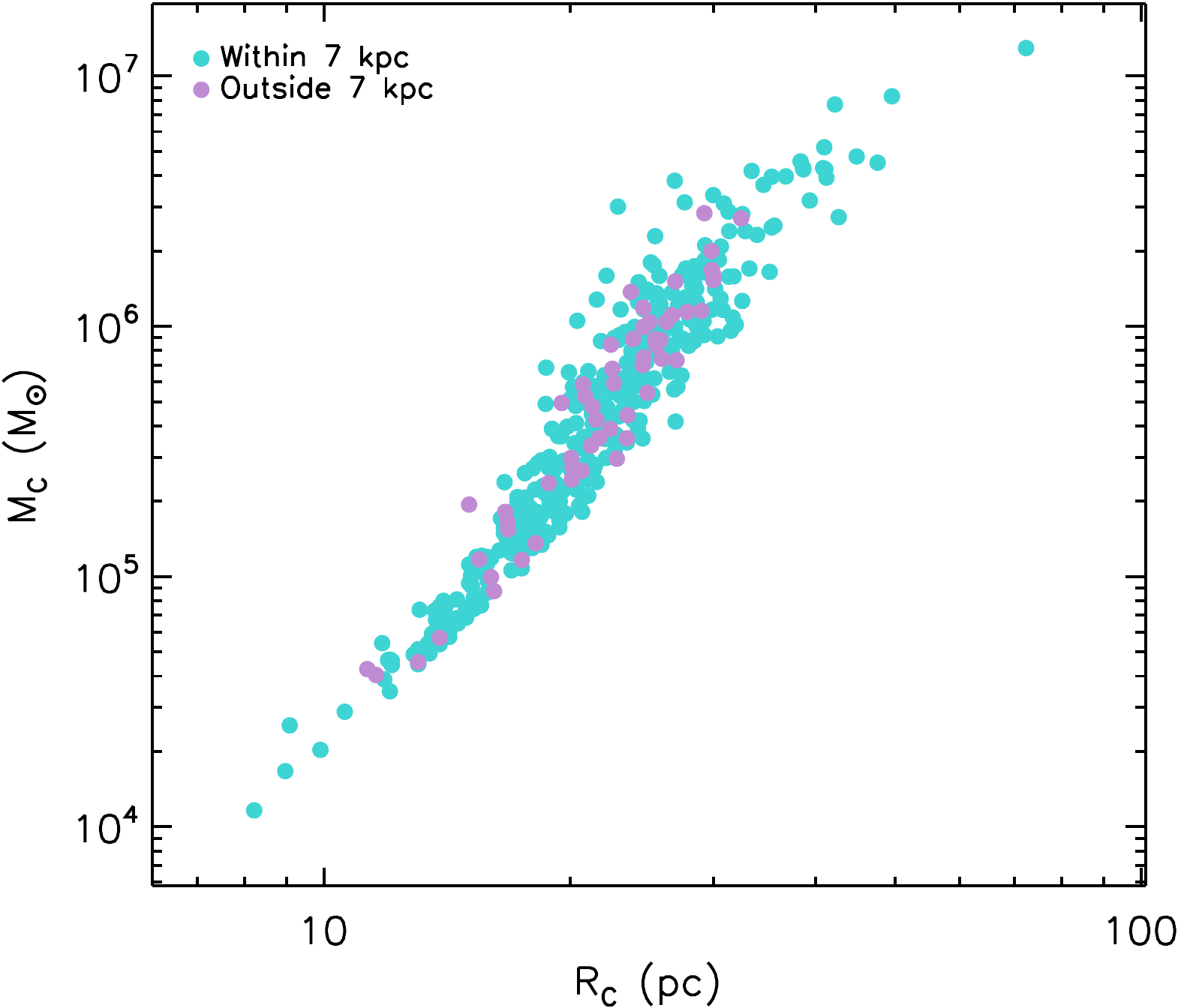}
\end{tabular}
\caption{Relationships for our cloud populations, divided into inner clouds (turquoise) and outer clouds (plum), left to right, top to bottom: velocity dispersion (\kms) versus radius (pc) compared to observations of the Milky Way \citep{Solomon1987} (dashed line) and the Local Groups \citep{Bolatto2008} (dash-dotted line); velocity dispersion divided by the radius's square root (\kms pc$^{-0.5}$) versus surface density (\msun pc$^{-2}$) with the line of unitary virial parameter (dashed line) compared to Milky Way data \citep{Heyer2009}; virial parameter versus cloud mass, with the line of unitary virial parameter (dashed line); and cloud mass (\msun) versus radius (pc).}
\label{fig:clships}
\end{figure*}

We present several relations in Fig. \ref{fig:clships} starting with the velocity dispersion versus cloud radius, first described by \citet{Larson1981}. We compare to the relation as found by \citet{Solomon1987} for the Milky Way and \citet{Bolatto2008} for the Local Group, which are quite similar. However, modern surveys reveal high scatter in this relation. \citet{Colombo2014} find an extremely weak correlation between velocity dispersion and radius and that most of their clouds exhibit a velocity dispersion above these two fits for every sector of M51. We also see a similar distribution as in M51, and a poor fit to this relation. This is due to our non-constant surface density for the cloud population.

More recently \citet{Heyer2009} suggest that instead of the Larson relation it is more useful to examine $\sigma_v / R_C^{0.5}$ versus $\Sigma_C$. Earlier surveys had a low spatial resolution and as a consequence mistakenly took the surface density to be constant across all clouds. Modern surveys show a range of surface densities across many galaxies \citep{Sun2018}, which we also find in Fig. \ref{fig:clhist}. The top right plot in Fig. \ref{fig:clships} shows our Heyer relation ($\sigma_v / R_C^{0.5}$ versus $\Sigma_C$) with the line of constant $\alpha_{vir}=1$. As in \citet{Heyer2009} most of our clouds are above this line and this relation is more linear for our data than the Larson relation above. 

In the bottom left plot in Fig. \ref{fig:clships}, we show the virial parameter versus cloud mass. 24 per cent of our clouds have $\alpha_{vir}<1$ and are unstable. Our average viral parameter is 2, which is in line with \citet{Rosolowsky2007}'s findings for M31, and that they argue is not significantly different from \citet{Solomon1987}'s value for the Milky Way, 1.45.

Lastly, the bottom right plot in Fig. \ref{fig:clships} gives the relation between cloud mass and radius. This plot is not compared to observation, however, but to another set of simulations. \citet{Fujimoto2016} in their simulations identify two separate cloud sequences that follow this relation differently, and also the same two sequences as in the Larson relation. In our simulations, however, we only find a single linear sequence. 

In all four plots in Fig. \ref{fig:clships} we also differentiate inner and outer clouds, and again find no difference between the two populations. 

\section{Discussion and Summary}
\label{sec:conclu}

\ramsesrt\ \citep{Rosdahl2013} is a moment-based radiative transfer expansion for the AMR hydrodynamics code \ramses\ \citep{Teyssier2002}. It produces, propagates, and destroys photons in  distinct groups related to the ionization of hydrogen and helium, and ties these photons and species to the thermal state of the gas.  \citet{Nickerson2018} presented the molecular hydrogen addition to \ramsesrt. The most novel aspect of our \htwo\ model was our self-shielding implementation. Previous codes modelled self-shielding by reducing the photodissociation of \htwo, whereas we modelled it by enhancing LW destruction. 

In this paper, we apply our molecular chemistry model to an isolated Milky Way-like disc galaxy at three different resolutions: Glow with a minimum cell width of 97 pc, GMed at 24 pc, and GHigh at 6.1 pc. This galaxy was also used in the AGORA code comparison project \citep {Kim2014,Kim2016}. We use the star formation model from \citet{Rasera2006}, which is based on the total gas density and is independent of \htwo. Furthermore, we choose not to enhance \htwo\ formation with a clumping factor. We seek to directly simulate the interaction of radiation and chemistry without any adjustable parameters on the galactic scale.


Just as with other simulations that maintain the star formation rate independent of \htwo\ density \citep{Capelo2018,Lupi2018}, the molecular-SFR relation observed by \citet{Bigiel2008} arises naturally. Tying \htwo\ to the star formation explicitly is not necessary to reproduce their relationship on the kpc scale. When considering the maximum \htwo\ fraction achieved, we only reach 0.985 in GHigh. This is a little lower than in other works \citep{Gnedin2011,Christensen2012,Hu2016,Capelo2018,Lupi2018} who do find volume elements with 100 per cent \htwo. These codes all use a clumping factor to account for unresolved cloud structure. However, our number is fairly close considering that we have no such factor. With sufficient resolution, a clumping factor may not be necessary.

There are several caveats to this work. We neglect magnetic fields, which play a role in molecular cloud dynamics \citep{Crutcher2012}, and a wider cosmological context. Inflows and outflows of gas from a galaxy can profoundly impact it. Dust, the main formation catalyst for \htwo\ and an important radiation shield, in not separately tracked in \ramsesrt. Instead we take the dust fraction of a cell to be the neutral hydrogen fraction. Stellar winds are another missing process. We are unable to resolve the St\"omgren spheres of ionized gas around hot, young stars, which we can fix through post-processing. We also cannot resolve lower mass clouds properly even at our highest resolution, GHigh. GLow does not converge well with the other two resolutions, but GMed does converge well with GHigh.

In summary, we preform a complete analysis of the \htwo\ gas in a Milky Way-like simulated disc galaxies from the morphology to individual molecular clouds, and find the effect resolution has on \htwo\ content. 

\textbf{Star formation history}: The star formation history of the galaxies helps us determine when all three resolutions reach a semi-steady equilibrium state. We choose 800 Myr as the time at which to analyze the galaxies in the subsequent sections.

\textbf{Morphology}: Resolution profoundly affects the structure and complexity of each galaxy. GLow is smooth and comparatively featureless aside from some spiral arms in the centre, while GMed has many more arms with clumps and GHigh not only has arms and clumps, but also intricate filamentary structure between the arms. \htwo\ traces the densest gas regions and remains confined to the disc, while \hi\ is more diffuse and additionally traces the gas ejected from the disc. \htwo\ effectively blocks the dissociating and ionizing radiation from young stars.

\textbf{Comparison to observables}: We present the observable properties of molecular gas on the kpc and overall gas content scales. Overall, our molecular fraction of neutral gas is comparable to observations for GHigh and GMed, but falls short for GLow. Our surface density profiles of \htwo, \hi, and SFR, like their morphology, change with resolution. GLow is dominated by \hi\ for its entirety, while GMed and GHigh host \htwo-dominated central regions. GLow and GMed have central SFR spikes and GHigh has a flat SFR profile, due to morphological quenching from its central bulge. We also compare the KS relation of our galaxies to data from a large survey \citep{Bigiel2008}. Our simulations fall within their margins, for both the total neutral gas relation and the pure molecular relation. We show, as in \citet{Schruba2011}, that the \htwo-SFR relation is tighter than the \hi-SFR relation.

\textbf{Phase diagrams}: These vivisect our galaxies into individual volume elements. $n_H$-$T$ diagrams show that all three galaxies largely follow the same contours, but how low the temperature goes and how high the density reaches depends on resolution. This profoundly affects the phase diagrams for \htwo\, while the other two hydrogen species are less affected by resolution. \htwo\ reaches peak abundances of 0.704, 0.925, and 0.985 in the GLow, GMed, and GHigh galaxies respectively.

\textbf{Molecular distribution}: We quantify the percentage of molecular gas in low-density regions. For GLow, GMed, and GHigh respectively half of the \htwo\ is in cells with densities below 5.57, 12.0, and 37.8 \cci\ and in cells with \htwo\ fractions below 0.254, 0.507, and 0.570. This shows that a significant amount of \htwo\ gas is diffuse and mixed with \hi.

\textbf{Molecular clouds}: We use GHigh to analyze our molecular clouds because it is the only resolution that produces gas concentrations similar to observed molecular clouds. Our molecular clouds trace the spiral arms, and the cumulative mass function is similar to that of the Milky Way. 25 per cent of \htwo\ is in these molecular clouds. None of the molecular clouds contain young stars. Our molecular clouds' properties (mass, radius, gas density, surface density, velocity dispersion, and the virial parameter) are similar to observations, and our average virial parameter is 2. However, when we split the histograms between the inner and outer disc, we get very similar distributions, unlike observations that show different distributions \citep{Miville-Deschenes2017}. Our clouds correspond to the \citet{Heyer2009} relation better than the \citet{Larson1981} relation, in keeping with modern findings that molecular clouds have non-constant surface densities. 

Considering the entire body of work between \citet{Nickerson2018} and this present paper, we have shown that the laws that govern \htwo\ on the chemical scale give rise to relations that govern \htwo\ on the galactic scale. We have demonstrated that high resolution is critical in galaxy simulations to properly form \htwo\ without the need for a clumping factor. 

In future, we can use our model further to explore both \htwo\ and \hi\ observations. It would be interesting to see how galaxy mergers affect both species \citep{Ellison2018}. On the molecular side, by adding CO to the code we can examine the relationship between \htwo\ and CO, the main observable of molecular gas, to further our work in Section \ref{ssec:dist}. Our model should be tested in regimes beyond a standard Milky Way disc, such as in low metallicity dwarf galaxies \citep{Hu2016}, or in the cold outflows from active galactic nuclei \citep{Aalto2015b}. Concerning \hi, we can now study the origins of the high velocity clouds above our own Milky Way and other galaxies \citep{Wakker1997}. On even grander scales, we can compare to large-scale \hi\ surveys such as the ALFALFA \citep{Haynes2011} in the context of the ``too big to fail problem''' \citep{Boylan-Kolchin2011} and results from the upcoming Square Kilometre Array \citep{Aharonian2013}. These are just a few of the many applications for \ramsesrt\ with both \htwo\ and \hi\ chemistry that we may explore in future.

\section*{Acknowledgements}
\label{sec:ack}
SN was supported by the University of Z\"urich Candoc Scholarship, and used the Piz Daint supercomputer in the Swiss National Supercomputing Centre in Lugano. JR was funded by the ORAGE project from the Agence Nationale de la Recherche under grant ANR-14-CE33-0016-03.



\bibliographystyle{mnras}
\bibliography{mendely.bib,sln_refs_ads.bib}







\bsp	
\label{lastpage}
\end{document}